\PassOptionsToPackage{unicode}{hyperref}
\PassOptionsToPackage{hyphens}{url}
\PassOptionsToPackage{dvipsnames,svgnames,x11names}{xcolor}
\documentclass[
  letterpaper,
]{article}

\usepackage{amsmath,amssymb}
\usepackage{iftex}
\ifPDFTeX
  \usepackage[T1]{fontenc}
  \usepackage[utf8]{inputenc}
  \usepackage{textcomp} 
\else 
  \usepackage{unicode-math}
  \defaultfontfeatures{Scale=MatchLowercase}
  \defaultfontfeatures[\rmfamily]{Ligatures=TeX,Scale=1}
\fi
\usepackage{lmodern}
\ifPDFTeX\else  
\fi
\IfFileExists{upquote.sty}{\usepackage{upquote}}{}
\IfFileExists{microtype.sty}{
  \usepackage[]{microtype}
  \UseMicrotypeSet[protrusion]{basicmath} 
}{}
\makeatletter
\@ifundefined{KOMAClassName}{
  \IfFileExists{parskip.sty}{%
    \usepackage{parskip}
  }{
    \setlength{\parindent}{0pt}
    \setlength{\parskip}{6pt plus 2pt minus 1pt}}
}{
  \KOMAoptions{parskip=half}}
\makeatother
\usepackage{xcolor}
\ifx\paragraph\undefined\else
  \let\oldparagraph\paragraph
  \renewcommand{\paragraph}[1]{\oldparagraph{#1}\mbox{}}
\fi
\ifx\subparagraph\undefined\else
  \let\oldsubparagraph\subparagraph
  \renewcommand{\subparagraph}[1]{\oldsubparagraph{#1}\mbox{}}
\fi

\providecommand{\tightlist}{%
  \setlength{\itemsep}{0pt}\setlength{\parskip}{0pt}}\usepackage{longtable,booktabs,array}
\usepackage{calc} 
\usepackage{etoolbox}
\makeatletter
\patchcmd\longtable{\par}{\if@noskipsec\mbox{}\fi\par}{}{}
\makeatother
\IfFileExists{footnotehyper.sty}{\usepackage{footnotehyper}}{\usepackage{footnote}}
\makesavenoteenv{longtable}
\usepackage{graphicx}
\makeatletter
\def\maxwidth{\ifdim\Gin@nat@width>\linewidth\linewidth\else\Gin@nat@width\fi}
\def\maxheight{\ifdim\Gin@nat@height>\textheight\textheight\else\Gin@nat@height\fi}
\makeatother
\setkeys{Gin}{width=\maxwidth,height=\maxheight,keepaspectratio}
\makeatletter
\def\fps@figure{htbp}
\makeatother
\NewDocumentCommand\citeproctext{}{}

\makeatletter
 \let\@cite@ofmt\@firstofone
 \def\@biblabel#1{}
 \def\@cite#1#2{{#1\if@tempswa , #2\fi}}
\makeatother
\newlength{\cslhangindent}
\newlength{\csllabelwidth}
\newenvironment{CSLReferences}[2] 
 {\begin{list}{}{%
  \setlength{\itemindent}{0pt}
  \setlength{\leftmargin}{0pt}
  \setlength{\parsep}{0pt}
  \ifodd #1
   \setlength{\leftmargin}{\cslhangindent}
   \setlength{\itemindent}{-1\cslhangindent}
  \fi
  \setlength{\itemsep}{#2\baselineskip}}}
 {\end{list}}
\usepackage{calc}

\newcommand{\CSLLeftMargin}[1]{\parbox[t]{\csllabelwidth}{\strut#1\strut}}
\newcommand{\CSLRightInline}[1]{\parbox[t]{\linewidth - \csllabelwidth}{\strut#1\strut}}

\usepackage{booktabs}
\usepackage{longtable}
\usepackage{array}
\usepackage{multirow}
\usepackage{wrapfig}
\usepackage{float}
\usepackage{colortbl}
\usepackage{pdflscape}
\usepackage{tabu}
\usepackage{threeparttable}
\usepackage{threeparttablex}
\usepackage[normalem]{ulem}
\usepackage{makecell}
\usepackage{xcolor}
\usepackage{siunitx}

  \newcolumntype{d}{S[
    input-open-uncertainty=,
    input-close-uncertainty=,
    parse-numbers = false,
    table-align-text-pre=false,
    table-align-text-post=false
  ]}
  
\usepackage[utf8]{inputenc}
\usepackage{todonotes}
\usepackage{multirow}
\usepackage{wrapfig}
\usepackage{float}
\usepackage{colortbl}
\usepackage{makecell}
\usepackage{xcolor}
\usepackage{siunitx}
\usepackage{orcidlink}
\usepackage{amsmath}
\usepackage[T1]{fontenc}
\usepackage{arxiv}
\makeatletter
\@ifpackageloaded{bookmark}{}{\usepackage{bookmark}}
\makeatother
\makeatletter
\@ifpackageloaded{caption}{}{\usepackage{caption}}
\AtBeginDocument{%
\ifdefined\contentsname
  \renewcommand*\contentsname{Table of contents}
\else
  \newcommand\contentsname{Table of contents}
\fi
\ifdefined\listfigurename
  \renewcommand*\listfigurename{List of Figures}
\else
  \newcommand\listfigurename{List of Figures}
\fi
\ifdefined\listtablename
  \renewcommand*\listtablename{List of Tables}
\else
  \newcommand\listtablename{List of Tables}
\fi
\ifdefined\figurename
  \renewcommand*\figurename{Figure}
\else
  \newcommand\figurename{Figure}
\fi
\ifdefined\tablename
  \renewcommand*\tablename{Table}
\else
  \newcommand\tablename{Table}
\fi
}
\@ifpackageloaded{float}{}{\usepackage{float}}
\floatstyle{ruled}
\@ifundefined{c@chapter}{\newfloat{codelisting}{h}{lop}}{\newfloat{codelisting}{h}{lop}[chapter]}
\floatname{codelisting}{Listing}

\makeatother
\makeatletter
\@ifpackageloaded{caption}{}{\usepackage{caption}}
\@ifpackageloaded{subcaption}{}{\usepackage{subcaption}}
\makeatother
\makeatletter
\@ifpackageloaded{algorithm}{}{\usepackage{algorithm}}
\makeatother
\makeatletter
\@ifpackageloaded{algpseudocode}{}{\usepackage{algpseudocode}}
\makeatother
\ifLuaTeX
  \usepackage{selnolig}  
\fi
\usepackage{bookmark}

\IfFileExists{xurl.sty}{\usepackage{xurl}}{} 
\hypersetup{
  pdftitle={Using Chao's Estimator as a Stopping Criterion for Technology-Assisted Review},
  pdfauthor={M.P. Bron; P.G.M. van der Heijden; A.J. Feelders; A.P.J.M. Siebes},
  colorlinks=true,
  linkcolor={blue},
  filecolor={Maroon},
  citecolor={Blue},
  urlcolor={Blue},
  pdfcreator={LaTeX via pandoc}}

\renewcommand{\today}{2024-03-30}
\title{Using Chao's Estimator as a Stopping Criterion for
Technology-Assisted Review}
\def\asep{\\\\\\ } 
\def\asep{\And }
\author{\textbf{M.P. Bron}~\orcidlink{0000-0002-4823-6085}\\\\Utrecht
University\\Utrecht\\\\The Netherlands National Police\\The
Hague\\\href{mailto:m.p.bron@uu.nl}{m.p.bron@uu.nl}\asep\textbf{P.G.M.
van der Heijden}~\orcidlink{0000-0002-3345-096X}\\\\Utrecht
University\\Utrecht\\\\University of
Southampton\\Southampton\\\href{mailto:p.g.m.vanderheijden@uu.nl}{p.g.m.vanderheijden@uu.nl}\asep\textbf{A.J.
Feelders}~\orcidlink{0000-0003-4525-1949}\\\\Utrecht
University\\Utrecht\\\href{mailto:a.j.feelders@uu.nl}{a.j.feelders@uu.nl}\asep\textbf{A.P.J.M.
Siebes}~\orcidlink{0000-0002-5108-7965}\\\\Utrecht
University\\Utrecht\\\href{mailto:a.p.j.m.siebes@uu.nl}{a.p.j.m.siebes@uu.nl}}
\date{2024-03-30}
\begin{document}
\maketitle
\begin{abstract}
Technology-Assisted Review (TAR) aims to reduce the human effort
required for screening processes such as abstract screening for
systematic literature reviews. Human reviewers label documents as
relevant or irrelevant during this process, while the system
incrementally updates a prediction model based on the reviewers'
previous decisions. After each model update, the system proposes new
documents it deems relevant, to prioritize relevant documents over
irrelevant ones. A stopping criterion is necessary to guide users in
stopping the review process to minimize the number of missed relevant
documents and the number of read irrelevant documents. In this paper, we
propose and evaluate a new ensemble-based Active Learning strategy and a
stopping criterion based on Chao's Population Size Estimator that
estimates the prevalence of relevant documents in the dataset. Our
simulation study demonstrates that this criterion performs well on
several datasets and is compared to other methods presented in the
literature.
\end{abstract}

\renewcommand{\Return}{\State \textbf{return}~}
\newcommand{\Print}{\State \textbf{print}~}
\newcommand{\Break}{\State \textbf{break}}
\newcommand{\Continue}{\State \textbf{continue}}
\newcommand{\True}{\textbf{true}}
\newcommand{\False}{\textbf{false}}
\renewcommand{\And}{\textbf{and}~}
\newcommand{\Or}{\textbf{or}~}
\newcommand{\Not}{\textbf{not}~}
\newcommand{\To}{\textbf{to}~}
\newcommand{\DownTo}{\textbf{downto}~}

\floatname{algorithm}{Algorithm}

\bookmarksetup{startatroot}

\section{Introduction}\label{introduction}

In extensive studies such as legal proceedings, criminal investigations,
and systematic reviews in academia, researchers and investigators gather
evidence or information by screening information found in large text
databases or corpora. The task is to find all pieces of information
relevant to the subject of the investigation. Often, the investigator
starts by using (Boolean) search queries to pre-select documents from
the database. Formulating these queries is not a trivial task, as it is
the objective to capture (nearly) all relevant documents. Often, the
resulting set of candidate documents that the researchers have to
process is enormous, while the prevalence of relevant documents within
these sets can be very low.

More formally, we have a dataset \(\mathcal{D}\) containing candidate
documents. During the review process, these documents are read by domain
experts and labeled as either \emph{relevant} or \emph{irrelevant}. Read
documents are referred to as \emph{labeled}, and we maintain two sets
\(\mathcal{L}^+\) and \(\mathcal{L}^-\), for the relevant and irrelevant
documents, respectively. The objective is to find all the remaining
\emph{unlabeled} relevant documents belonging to the set
\(\mathcal{U}^{+}\).

The prevalence for systematic review corpora ranges from below 1 to 35
\% {[}8{]}, so most candidate documents are not relevant. Traditionally,
the investigator reviewed each document in \(\mathcal{D}\), resulting in
a large amount of work. Recently, systems were proposed and built that
potentially reduce the human effort needed by limiting the number of
irrelevant documents shown to the reviewer {[}9, 15, 37, 43{]}. These
systems use machine learning to recommend documents based on prior
review decisions. We refer to these systems as \emph{Technology-Assisted
Review} (TAR) systems as coined in {[}15{]}. Many recent TAR systems use
\emph{Active Learning} (AL) to update the classifier after each or
several review decisions iteratively. AL is a machine learning technique
that is used to train a classifier with fewer labeled data points while
retaining good performance. In this setting, the model can interactively
query an oracle (i.e., the domain expert) to label data points with the
desired output of the Machine Learning model (i.e., in the case of a
classification task, the class of the data point). In our case, the
model should predict the relevancy of each document. Because the model
is frequently retrained, it could reduce the number of instances that
should be labeled by querying the most informative documents. Many TAR
systems that use AL show the user the top-\(k\) unseen documents
according to the classifier's predictions. As the classifier is
retrained frequently, the ranking is refined as well, reducing the
number of documents that have to be screened.

In the case of abstract screening for systematic reviews,
state-of-the-art systems can find all relevant documents after screening
only 5 -- 40 \% of the corpus by using this general methodology {[}9,
37{]}. A caveat is that these systems lack reliable stopping criteria.
Simulation studies show that we can reduce work if we know the
prevalence \emph{a priori}. However, the number of relevant documents is
not known beforehand in a real-world situation. Because of this, an
investigator may stop too early, resulting in the omission of important
information. Conversely, stopping too late causes unnecessary effort.

In this work, we describe a method to determine the prevalence of
relevant documents using \emph{Population Size Estimation} (PSE)
methods. These methods are used in official statistics and public health
to estimate population size when only part of the population is
observed. PSE methods are related to \emph{Capture Mark Recapture
models}, originating from ecology, where these models are used to
estimate the population size of wildlife. In our case, we want to
estimate the size of the set of relevant documents, i.e., the number of
relevant documents. During systematic reviews, only a subset of the
relevant documents is observed, that is, only the set of documents that
the investigators read. A review can only be stopped if the reviewers
believe no relevant documents have been missed or their recall target is
met.

In this work, we investigate if PSE is a suitable technique for deciding
when to terminate the TAR procedure. As our main contribution, we show
how two versions of Chao's Estimator {[}10{]}, a PSE method, can be
integrated into a TAR system and used within a stopping criterion for
the review process. Furthermore, we present the results of an extensive
simulation study in which we compared this stopping criterion to various
other methods presented in the literature.

\bookmarksetup{startatroot}

\section{Related Work}\label{related-work}

The task of Technology Assisted Review (TAR) systems is to retrieve a
significant number, if not all, of the relevant documents within a
dataset \(\mathcal{D}\). To achieve this, Active Learning is often
continuously applied to the dataset, a process commonly known as
Continuous Active Learning (CAL) in the literature. Continuous Active
Learning aims to minimize the number of irrelevant documents while
maximizing the number of retrieved relevant documents. Over the years,
Cormack and Grossman have developed a variety of CAL methods, with the
most prominent method being AutoTAR {[}14{]}. We describe the CAL
procedure in  Algorithm~\ref{alg-cal} .

\begin{algorithm}[htb!]
\caption{The Continuous Active Learning algorithm. 
The algorithm requires as parameters a dataset $\mathcal{D}$, an unlabeled set of documents $\mathcal{U}$, 
labeled documents $\mathcal{L}^+, \mathcal{L}^-$, a classifier $C$, a batch size $b$.
The Active Learning procedure selects new documents according to the relevance predictions of the classifier $C$, which are updated after each batch of labeling decisions.}
\label{alg-cal}
\begin{algorithmic}[1]
\Procedure{CAL}{$\mathcal{D}, \mathcal{U}, \mathcal{L}^+, \mathcal{L}^-, C, b$}
  \State $S \gets$ \False \Comment{Variable indicating whether CAL can be stopped}
  \While{$|\mathcal{U}| > 0$ \And \Not $S$}
    \State $C$\Call{.Fit}{$\mathcal{L}^+, \mathcal{L}^-$}
    \State $\mathcal{B} \gets $ \Call{Select}{$\mathcal{U}, C, b$}
    \For{$d \in \mathcal{B}$}
      \State $y \gets $ \Call{Review}{$d$} \Comment{Performed by the human reviewer}
      \If{$y =$ Relevant}
        \State $\mathcal{L}^+ \gets \mathcal{L}^+ \cup \{ d \} $ 
      \Else
        \State $\mathcal{L}^- \gets \mathcal{L}^- \cup \{ d \} $ 
      \EndIf
      \State $\mathcal{U} \gets \mathcal{U} \setminus \{ d \} $ 
    \EndFor
    \State $S \gets$ \Call{StoppingCriterion}{$\mathcal{D}, \mathcal{U}, \mathcal{L}^+, \mathcal{L}^-, C, b$}
  \EndWhile
  \State \textbf{return} $\mathcal{L}^+, \mathcal{L}^-$
\EndProcedure

\Procedure{Select}{$\mathcal{U}, C, b$}
  \State $\mathbf{P} \gets C$\Call{.Predict}{$\mathcal{U}$} \Comment{Returns the relevance score for all $d$ in $\mathcal{U}$}
  \State $\mathbf{R} \gets $\Call{Rank}{$\mathcal{U}, \mathbf{P}$}
  \State $\mathcal{B} \gets $ \Call{Head}{$\mathbf{R}, \mathcal{U}, b$} \Comment{Gets the top-$b$ documents}
  \State \textbf{return} $\mathcal{B}$
\EndProcedure
\end{algorithmic}
\end{algorithm}

Many CAL procedures require a set of seed documents provided by the
reviewer. This set needs to contain at least one relevant document, but
it does not need to be a document from \(\mathcal{D}\); it may also
contain a topic description. Additionally, one example of an irrelevant
document is needed. These are then used as inputs \(\mathcal{L}^+\) and
\(\mathcal{L}^-\) for  Algorithm~\ref{alg-cal} . In each iteration, a
classifier is fitted to the currently labeled information. Then, a batch
containing the top-\(b\) documents is selected according to the ranking
based on the classifier's predictions. After the user has labeled each
document in the batch, the process is repeated until \(\mathcal{U}\) is
empty or a stopping criterion has been triggered. The procedure aims to
optimize the retrieval of relevant documents by updating the classifier
each iteration. Given a good stopping criterion, this procedure enables
the user to minimize the workload while finding nearly all relevant
documents.

AutoTAR is an adaptation of the CAL procedure {[}14{]}. Instead of just
training on the labeled documents \(\mathcal{L}^+, \mathcal{L}^-\), it
samples a set of documents from the unlabeled set \(\mathcal{U}\), which
are temporarily assumed to be irrelevant. This is a fair assumption,
given the low prevalence of relevant documents in most datasets.
Moreover, AutoTAR increases the batch size of each iteration by 10 \%.
ASReview {[}37{]} has a fixed batch size of 1 and uses dynamic
resampling to deal with imbalanced training data to improve the
classifier's performance. In recent years, several other algorithms that
also adhere to the CAL paradigm have been proposed, with each their own
adjustments, have been proposed (inter alia {[}9, 43{]}).

\subsection{Stopping Criteria}\label{sec-stopping-criteria-relatedwork}

As described above, the CAL procedure leaves the question open of how to
stop the review process (i.e., the \textsc{StoppingCriterion} procedure,
line 15 in  Algorithm~\ref{alg-cal} , is not given). Researchers have
recently developed various approaches to solve the stopping problem. In
{[}30{]}, the authors provide a taxonomy to classify the diverse range
of stopping criteria. The authors classify the criteria according to two
axes, namely \emph{applicability to TAR methods} and the
\emph{guarantees} these methods offer. For applicability, each method
can fall into one of the following three categories.

\begin{description}
\item[Interventional methods.]
This category of rules intervenes in the selection strategy of
documents. Some interventional methods depend on a specific sampling
methodology; others even deviate from the general CAL paradigm. These
alterations enable the usage of specific statistical methods or tests.
Some sampling strategies allow the use of an estimator to estimate the
number of relevant documents within the corpus.
\item[Standoff methods.]
Methods that fall in this category can be used in combination with any
TAR system, as these methods do not depend on any sampling strategy.
\item[Hybrid methods.]
Some methods interleave or divide the process into phases, alternating
the original method with periods in which another sampling strategy is
used.
\end{description}

For the second axis, \emph{guarantees}, each method falls into one of
the following two categories.

\begin{description}
\item[Heuristic.]
Heuristic rules make a stopping decision based on general patterns
observed in, for example, the recall statistics of the review. However,
as these methods do not have a formal statistical grounding, they do not
offer strong guarantees besides the results of the criterion on known
datasets.
\item[Certification rules.]
Certification rules provide a formal statistical guarantee that the
stopping point has certain properties and/or that the rule provides a
formal statistical estimate of effectiveness at the stopping point.
\end{description}

In the following sections, we list several criteria and list their
classification according to this taxonomy.

\subsubsection{Pragmatic Criteria (Standoff \&
Heuristics)}\label{pragmatic-criteria-standoff-heuristics}

Pragmatic criteria are often based on the recall statistics of the
process. A commonly used heuristic is to stop the TAR after \(k\)
consecutive irrelevant document suggestions. Examples of values of \(k\)
found in the literature are 50 and 200 {[}8{]}. However, studies have
shown that this method often results in low recall or little work
savings {[}8{]}. Moreover, this method frequently fails to meet the
widely used recall target of 95\%.

Another trivial method is to stop screening after reviewing half of the
documents in \(\mathcal{D}\) {[}39{]}. A variant based on this is the
criterion Rule2399 {[}16{]}, which stops the procedure if the size of
the set of read documents satisfies
\(|\mathcal{L}| \geq 1.2 \cdot |\mathcal{L}^+| + 2399\).

\subsubsection{Baseline Inclusion Rate (Hybrid \&
Heuristic)}\label{baseline-inclusion-rate-hybrid-heuristic}

An example of a hybrid method is the Baseline Inclusion Rate {[}35{]}.
In this approach, a random sample \(\mathcal{S}\) of the dataset
\(\mathcal{D}\) is taken initially, before the TAR procedure. All the
documents in \(\mathcal{S}\) are then reviewed. If \(\mathcal{S}\) is
large enough, the ratio \(\frac{|\mathcal{S}^+|}{|\mathcal{S}|}\) should
approximate the ratio \(\frac{|\mathcal{D}^+|}{|\mathcal{D}|}\), where
\(\mathcal{S}^+\) and \(\mathcal{D}^+\) denote the subsets of relevant
documents in \(\mathcal{S}\) and \(\mathcal{D}\), respectively. After
labeling \(\mathcal{S}\), the Active Learning phase is started. The
process is then stopped when
\(|\mathcal{L}^+| \geq \frac{|\mathcal{S}^+|}{|\mathcal{S}|}|\mathcal{D}|\).
However, due to sampling uncertainty, a large sample may be needed to
obtain a good estimate of the prevalence. This process may consume a lot
of time, and since the sample is random, there is no guarantee of saving
any work during this period. Furthermore, this estimation is static,
which may lead to a review with no work savings if the estimate is even
slightly too high {[}8{]}.

\subsubsection{Target method (Hybrid \&
Heuristic)}\label{target-method-hybrid-heuristic}

In {[}15{]}, the Target method is proposed, which aims for high recall
and guarantees a recall of at least 70 \%. This method divides the TAR
process into two phases. First, the method randomly samples documents
from the dataset until \(k\) relevant documents are found. The size of
\(k\) depends on the user and the dataset, but the recommended value for
\(k=10\), according to {[}15{]}. When these documents have been found,
the system proceeds to the second phase by starting a standard TAR
procedure, for example, AutoTAR. However, the documents' judgments from
the previous phase are not given to the TAR system of the second phase,
so, from the machine learning perspective, the TAR procedure restarts
from scratch. The stopping criterion is triggered when all \(k\)
relevant documents from the first phase are rediscovered during the
second phase.

\subsubsection{Knee method (Standoff \&
Heuristic)}\label{knee-method-standoff-heuristic}

An already established heuristic is the \emph{knee method} {[}15{]}.
Most recall curves from TAR systems have an inflection point (which
looks like a knee, hence the name). This method compares the slopes
before and after the knee. When the ratio \(\rho\) between the two
slopes becomes larger than a specific threshold or bound, the review
process should be stopped. The slope ratio can be calculated as follows
{[}15, 41{]}:

\[
\rho(\mathcal{L}_t) = \frac{|\mathcal{L}^+_i|}{|\mathcal{L}_t|} \frac{
|\mathcal{L}_t|
-|\mathcal{L}_i|}{|\mathcal{L}^+_t|-|\mathcal{L}^+_i| + 1}\ ,
\]

\noindent where \(t\) is the current iteration and \(i\) is the
iteration that maximizes the perpendicular distance between the point
\((|\mathcal{L}_i|,|\mathcal{L}^+_i|)\) and the line that goes through
the origin \((0,0)\) and the point \((t, |\mathcal{L}^+_t|)\). The bound
is dynamic; it decreases as the number of relevant documents increases.
In {[}15{]}, the bound for iteration \(t\) is defined as
\(\textrm{bound}_t = 156 - \min (|\mathcal{L}^+_t|, 150)\). The Knee
criterion is triggered when
\(\rho(\mathcal{L}_t) \geq \textrm{bound}_t\) and
\(|\mathcal{L}_t| \geq 1000\). The Knee method is designed with the
batch size scheme of AutoTAR in mind. However, this method can easily be
adjusted to work with any batching scheme {[}30{]}, so this method is a
standoff method.

\subsubsection{Budget method (Standoff \&
Heuristic)}\label{budget-method-standoff-heuristic}

The Budget method {[}15{]} combines aspects of both the Knee method and
the Target method, as well as observations on the recall statistics of
TAR systems. This method can be stopped when either of the following two
criteria are met:

\begin{enumerate}
\def\labelenumi{\arabic{enumi}.}
\item
  The first criterion is based on the observation that after reading
  75\% of the dataset \(\mathcal{D}\) using random sampling, we can
  assume that we have found approximately 75\% of the relevant
  documents. Additionally, we can assume that most TAR methods will
  improve upon random sampling. Therefore, the Budget method specifies
  that we can stop when the size of the read documents
  \(|\mathcal{L}| \geq 0.75|\mathcal{D}|\).
\item
  The second criterion is based on the Knee method and Target method. We
  can observe that during phase one of the Target method, with target
  set size \(k\), the number of randomly sampled documents would be
  \(|\mathcal{L}_{\text{target}}| = k \frac{|\mathcal{D}|}{|\mathcal{D}^+|}\)
  for a dataset \(\mathcal{D}\) and its positive component
  \(\mathcal{D}^+\). At each iteration, we record the set of relevant
  documents as \(\mathcal{L}^+\). Since
  \(\mathcal{L}^+ \subseteq \mathcal{D}^+\),
  \(k\frac{|\mathcal{D}|}{|\mathcal{L}^+|}\) should be at least as large
  as the random sample size in the Target method. Combined with the
  slope ratio of the Knee method, the Budget method is triggered when
  \(\rho(\mathcal{L}) \geq 6\) and
  \(\mathcal{L} \geq k\frac{|\mathcal{D}|}{|\mathcal{L}^+|}\).
\end{enumerate}

Just like the Knee method, the Budget method is designed with the batch
size scheme of AutoTAR in mind, but can be adapted easily to work with
any batching scheme {[}30{]}.

\subsubsection{AutoStop (Interventional \&
Certification)}\label{sec-rl-autostop}

Whereas the previously discussed methods are heuristics, the AutoStop
method {[}31{]} is a topic-wise interventional certification method and
is thus tightly coupled to its sampling strategy. The method aims to
estimate the number of relevant documents \(\mathcal{D}^+\) and use that
estimate to decide when to stop by calculating the expected recall.
AutoStop consists of four modules.

\begin{enumerate}
\def\labelenumi{\arabic{enumi}.}
\item
  \emph{Ranking module}. The procedure is similar to AutoTAR from the
  machine learning perspective. However, the main difference is on the
  inference side: the trained model is used to process all documents in
  \(\mathcal{D}\) instead of only the unlabeled set \(\mathcal{U}\). The
  resulting posterior probabilities are then used to produce a ranking.
\item
  \emph{Sampling module}. The sampling strategy is unique in that it
  makes major adjustments to the sampling procedure of the CAL procedure
  (see  Algorithm~\ref{alg-cal} ). The normal CAL procedure (as in
   Algorithm~\ref{alg-cal} ) selects a batch with the top-\(k\)
  documents in \(\mathcal{U}\). Instead, AutoStop makes a ranking on
  \(\mathcal{D}\). Then, this ranking is used to sample, with
  replacement, from \(\mathcal{D}\) where each document \(d\) is
  weighted according to its rank so higher-ranked documents have a
  higher sampling probability.
\item
  \emph{Estimation module}. This sampling strategy enables us to produce
  an unbiased estimate of the total number of relevant documents from
  the sampling history. An estimate can be calculated using either the
  Horvitz-Thompson estimator or Hansen-Hurwitz estimator. Besides a
  point estimate, we can also calculate its variance and, subsequently,
  a confidence interval of the estimate.
\item
  \emph{Stopping module}. The stopping module offers two strategies: an
  \emph{optimistic} stopping criterion calculates the expected recall
  according to the point estimate. The stopping criterion is triggered
  when the recall target that the user has set has been achieved. The
  other is a \emph{conservative} criterion, which instead bases its
  decision on the upper bound of the estimate's CI.
\end{enumerate}

The estimation module estimates the size of \(\mathcal{D}^+\) and the
variance of this estimate. This feature allows the creation of stopping
criteria with several recall targets and confidence levels. A downside
of this method is its memory usage. The estimation module consumes
approximately 20 GB for a set of 15000 documents, and memory consumption
grows quadratically in terms of the dataset size {[}31{]}. Larger
datasets must be divided into smaller manageable parts, and then the
AutoStop procedure is performed for each part separately to overcome
memory limitations. Unfortunately, this creates some additional overhead
as knowledge is not shared between parts.

\subsubsection{Quant (CI) Rule (Standoff \&
Certification)}\label{quant-ci-rule-standoff-certification}

The Quant rule {[}41{]} bases its estimate on the relevance
probabilities of labeled documents and the unlabeled documents. Assuming
that the model is well calibrated, that is, supposing we take a large
sample of documents with a given probability \(p\), then the prevalence
of relevant documents is approximately \(p\). Suppose at iteration \(t\)
we have fitted a model with parameters \(\theta_t\), then we can
estimate the number of relevant documents in the set of labeled
documents as follows. \[
\widehat{|{\mathcal{L}^+_t}|} = \sum_{j \in \mathcal{L}_t} p\left( y=1|d_j; \theta_t \right) \quad .
\] For the unlabeled documents, a similar procedure is performed: \[
\widehat{|{\mathcal{U}^+_t}|}= \sum_{j \in \mathcal{U}_t} p\left(y=1|d_j; \theta_t \right) \quad .
\] Then, the recall can be estimated as follows: \[
\hat{R}_t = \frac{\widehat{|{\mathcal{L}_t}|}}{\widehat{|{\mathcal{L}_t}|} + \widehat{|{\mathcal{U}_t}|}} \quad .
\] When \(\hat{R}_t \geq R_{\text{tar}}\), where \(R_\textrm{tar}\)
denotes the target recall, then the stopping criterion is triggered.
Besides this point estimate, {[}41{]} provide a method to calculate the
variance of this estimator, which in turn can be used to calculate a 95
\% confidence interval (\(\pm\) 2 standard deviations). Given the size
of \(\mathcal{L}^+\), the recall estimate \(\hat{R}\) can be used to
produce an estimate of the size of \(\widehat{|\mathcal{D}^+|}\). These
estimates can then be used for a conservative and optimistic stopping
criterion in a similar fashion as in AutoStop.

\subsubsection{Hypergeometric method (Hybrid, Standoff \&
Certification)}\label{sec-rl-cmh}

In {[}8{]}, a statistical stopping criterion for abstract screening for
systematic reviews based on statistical testing is introduced. In this
work, both a hybrid and standoff method are proposed. Their method is
centered around the hypergeometric distribution. The authors assert that
the number of missing relevant papers contained in a random sample
follows the hypergeometric distribution.

The standoff version of their method works as follows. After each
iteration, the labeled set is divided into two parts around a pivot
iteration \(i\). The method then calculates the probability that the
current recall equals or exceeds the target recall. The recall target is
specified as \(\tau_{\text{tar}}\). Then, we iterate over each pivot
\(i\) and calculate the probability

\[
p_i = P \left( X \leq \mathcal{L}^+ - \mathcal{L}^+_i \right)\quad ,
\]where

\[X \sim \textrm{Hypergeometric}(\mathcal{D} - \mathcal{L}^+_i, K_{\textrm{tar}}, \mathcal{L} - \mathcal{L}_i)\quad ,\]
and \[
K_{\textrm{tar}} = \Biggl \lfloor \frac{\mathcal{L}^+}{\tau_{\textrm{tar}}} - \mathcal{L}^+_i + 1 \Biggr \rfloor \quad .
\]After iterating over all pivots, the TAR process is stopped if there
is an iteration \(i\) where \(\min(p_i) < \alpha\) for some confidence
level \(\alpha\). In {[}8{]} \(\alpha = 0.05\) is chosen. We will refer
to this method as \emph{CMH-Standoff}.

Their hybrid (Ranked Quasi Sampling strategy in {[}8{]}) method consists
of two phases. In the first phase, the TAR procedure is followed as
normal until the standoff criterion, as described above, is triggered
(here \(\alpha = 0.5\)). Random sampling will then be used for the
remaining part of the screening. Like the standoff version, the test is
repeated each iteration; however, with one alteration, the pivot
iteration \(i\) is fixed on the iteration in which the standoff rule was
triggered. This version only misses the target recall of 95 \% or above
in a few scenarios ({[}8{]} reports 0.59 \% of the runs on several
datasets with different seed documents). However, this robustness comes
at a cost, as their method relies on random sampling. The result is that
the average work reduction over random sampling (WSS, see
Equation~\ref{eq-wss}) achieved with their stopping criterion is only 17
\%. We will refer to this method as \emph{CMH-Hybrid}.

\bookmarksetup{startatroot}

\section{Methodology}\label{sec-methodology}

In our work, we propose a novel stopping criterion for
Technology-Assisted Review (TAR) that uses a Population Size Estimator
to estimate the size of the set of relevant documents \(\mathcal{D}^+\).
To be more precise, we adopt Chao's moment estimator {[}10{]} and a
Poisson Regression version of this estimator {[}32{]}. The sampling
procedure and estimator are intertwined, that is, without this specific
sampling procedure, these estimators cannot be used. First, we describe
Population Size estimators in the context of systematic search tasks and
TAR. Then, we describe our sampling procedure, followed by an overview
of the aforementioned estimators. This is followed by an overview of the
stopping criteria that use the estimates. This section concludes with a
more detailed description of the classification algorithms and the
Active Learning procedure.

\subsection{Population Size Estimation for Technology-Assisted
Review}\label{population-size-estimation-for-technology-assisted-review}

Population Size Estimation (PSE) techniques are commonly used to
estimate the total size of only partially observed populations, such as
animal and human populations {[}2{]}. Besides calculating the size of
human and animal populations, PSE methods were also applied to estimate
the number of other partially observed sets of objects or phenomena,
such as the number of hidden faults within a software package {[}13{]}.
PSE may involve linking multiple lists recording observations of
individuals or the number of times an individual is observed . These
records can then be used to determine the capture probabilities of
individuals, which in turn can be used to estimate the size of the
entire population. In the case of TAR, the estimand is the number of
relevant documents within a dataset, that is, the size of
\(\mathcal{D}^+\). The set \(\mathcal{D}^+\) consists of two parts, the
set that has been found by the user \((\mathcal{L}^+)\) and the set of
documents that have yet eluded the search process \((\mathcal{U}^+)\).
More formally, \(\mathcal{D}^+ = \mathcal{L}^+ \cup \mathcal{U}^+\). The
user can stop the process once the estimate
\(\widehat{|\mathcal{D}^+|}\) approaches \(|\mathcal{L}^+|\) and
consequently the \(\widehat{|\mathcal{U}^+|}\) becomes low enough.

\subsubsection{PSE for Search Tasks}\label{pse-for-search-tasks}

The use of PSE techniques for search tasks has been explored previously
in the literature {[}27, 33, 36, 40{]}. For example, {[}40{]} presented
a method employing a PSE to estimate the number of omissions from a
systematic literature review. The basic outline of this approach is as
follows: multiple reviewers conduct independent searches for documents
relevant to a specific topic and decide for each document they review if
it is relevant to this topic. For simplicity, we assume that the
reviewers are unanimous in deciding the relevancy of each encountered
document \(i\). The sets of relevant documents \(\mathcal{L}^+_j\) for
each reviewer \(j\) may differ, as the search skills of individual
reviewers vary. In the end, upon completing their tasks, the reviewers
link their sets \(\mathcal{L}^+_j\), to identify for each document \(i\)
the reviewers by which it was discovered.

The linking procedure is performed as follows (we are using the notation
from {[}10{]}). Suppose we have a review committee \(\mathcal{C}\)
consisting of \(C = \left| \mathcal{C} \right|\) reviewers. We represent
the result of the search process as a \(N \times C\) matrix
\(\mathbf{X} = (X_{ij})\) where \(N\) is the size of the set of
\emph{all} relevant documents (that is, both the documents that were
found and the documents that eluded the reviewers) and
\(C = \left| \mathcal{C} \right|\) is the size of the committee. Then,
we specify the elements of \(\mathbf{X}\) as
\[ X_{ij} = I\left[\text{document}\ i\ \text{is present in}\ \mathcal{L}^+\ \text{of reviewer}\  \mathcal{C}_j\right]\ ,
\]where \(I[A]\) is an indicator function: \(I[A] = 1\) if the event
\(A\) occurs and \(0\) otherwise. This results in a matrix in which each
row represents a document and each column a reviewer. The cells then
contain a 1 if the document \(i\) has been found by reviewer
\(\mathcal{C}_j\) and 0 otherwise. Then, \[
n = \sum^{N}_{i=1} I\left[ \sum^{C}_{j=1} X_{ij} \geq 1 \right]\ ,
\] denotes the number of distinct relevant documents that have been
found by at least one reviewer, in other words, the number of relevant
documents that has been found by the committee as a whole. Furthermore,
we specify the frequency statistic

\[
f_k = \sum^N_{i=1} I\left[ \sum^C_{j=1} X_{ij} = k\right],
\quad k = 0,\ 1,\ \dots,\ C,
\] which denotes the number of documents that have been found by exactly
\(k\) reviewers. Of course, matrix \(\mathbf{X}\) is not fully observed:
only the \(n\) rows for the documents that were found at least once are
observed and as we do not know how many documents have been missed by
all reviewers, we do not know the size of the \(N\) dimension of matrix
\(\mathbf{X}\). Consequently, we do not know how many documents have
frequency statistic \(f_0\). Then, the estimand can be defined as \[
\hat{N} = n + \hat{f_0}\ ,
\]where \(\hat{N}\) denotes the estimate for the total number of
relevant documents, of which \(\hat{f}_0\) are unobserved. In
Section~\ref{sec-chaos-estimator}, we will further discuss the models.

\subsubsection{PSE without multiple
reviewers}\label{sec-pse-without-multiple-reviewers}

A limitation of the approach sketched above is the need for multiple
human reviewers in the review procedure to enable the estimation of the
number of omitted relevant documents. In this work, we propose to adapt
the sampling strategy to allow us to estimate using PSE without relying
on multiple reviewers. We employ an ensemble of Active Learning methods
that individually rank and propose documents. For each method, we
maintain a registration list containing the identifiers of documents
identified by each method. Our approach draws on the Query-By-Committee
and Query-By-Bagging paradigms used in Active Learning, as introduced by
{[}34{]}. In Query-By-Committee, an ensemble
\(\mathcal{C} = { \mathcal{C}_1, \ldots, \mathcal{C}_n}\) is
constructed, comprising multiple classifiers of different classification
algorithms, such as Multinomial Naïve Bayes, Logistic Regression, and
Random Forest. Since each classifier has a distinct decision function,
each will likely produce a unique ranking. In Query-By-Bagging, each
classifier is presented with a unique subset of the labeled corpus. Our
approach combines both methods; we incorporate a diverse range of
classification algorithms and independent training sets for each
classifier.

The canonical Query-By-Committee method typically pools classifier
decisions using a query strategy such as \emph{Vote Entropy}. This
method involves each classifier in the ensemble voting for its
prediction on an unlabeled instance, and the instance with the most
disagreement among committee members is selected for human review.
However, in our approach, each committee member has its own query
strategy and functions as an independent TAR system. We do not use an
overarching query strategy that combines the results of these methods;
instead, each proposed document is selected by choosing one of the
members in a round-robin or random fashion. The selected committee
member then presents an instance for review according to its individual
query strategy.

\begin{figure}

\centering{

\includegraphics[width=1\textwidth,height=\textheight]{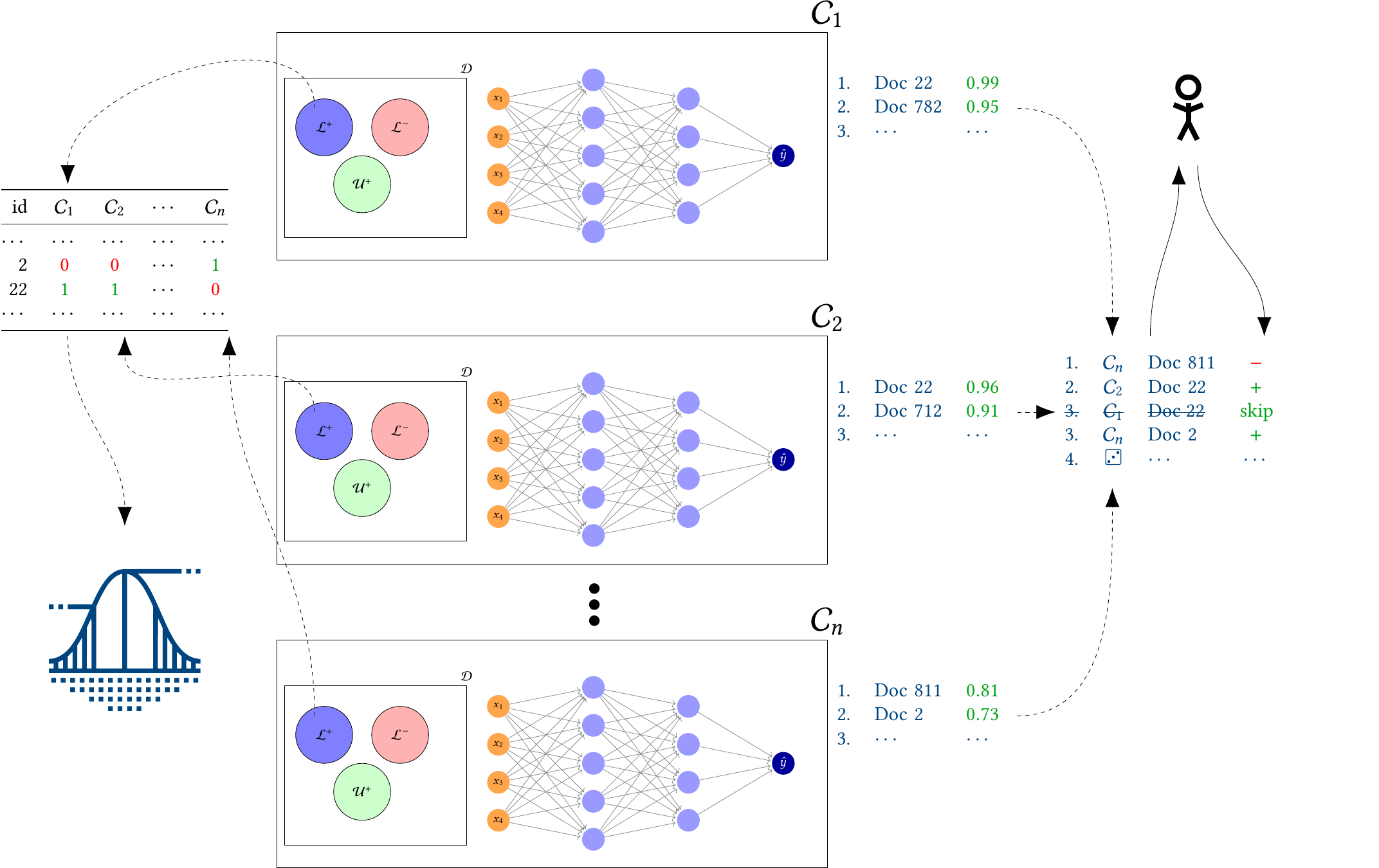}

}

\caption{\label{fig-architecture}This figure shows an architectural
overview of our method. The Active Learning module consists of several
committee members \(\{\mathcal{C}_1, \dots, \mathcal{C}_n \}\), with
each its own labeled and unlabeled state. Each of the members can have a
Machine Learning Model (for illustrative purposes represented as an
Artificial Neural Network). The rankings of each of the members are
combined by going through each member in a round-robin or random fashion
and selecting the top of the stack. The estimation module can query the
labeled states of each of the member to construct a contingency table
and fit a PSE model.}

\end{figure}

In this approach, it is possible for a member \(\mathcal{C}_i\) to
propose an instance \(d_k\) for review in iteration \(t\), which was
already proposed by \(\mathcal{C}_j\) in a previous iteration
\(t^\prime\). To handle this situation, our method ensures that the
label for \(d_k\) is given to the member \(\mathcal{C}_i\). This is
illustrated in Figure~\ref{fig-architecture}, where document 22 is first
proposed by \(\mathcal{C}_2\) and then in the next iteration by
\(\mathcal{C}_1\). The labeling decision is transferred to
\(\mathcal{C}_2\), and from the user's perspective, there is no
difference. The process continues by selecting one of the committee
members again until a document is proposed that has not been proposed by
any of the other methods.

After each label decision, the estimation module generates a matrix
\(\mathbf{X}\) from the system as follows.

\[ X_{ij} = I\left[\text{document}\ i\ \text{is present in}\ \mathcal{L}^+\ \text{from committee member}\  \mathcal{C}_j\right]\ .
\] \noindent Then, a PSE model uses \(\mathbf{X}\) to estimate the
number of omissions. The stopping module compares the estimate and its
confidence interval to the current recall statistics and decides if the
TAR procedure can be terminated. If not, this procedure is repeated by
retraining the models and sampling new documents from the updated
rankings.

\subsection{Chao's Moment Estimator}\label{sec-chaos-estimator}

In our work, we use Chao's moment estimator {[}10{]} and a Poisson
Regression adaptation by Rivest and Baillargeon {[}32{]}. While several
PSE methods use the full matrix \(\mathbf{X}\) to model \(\hat{N}\),
both models only use the frequency statistics \(f_k\). In the following
sections, we will introduce Chao's estimator and the Poisson regression
adaptation through an example.

\begin{figure}

\centering{

\includegraphics[width=0.75\textwidth,height=\textheight]{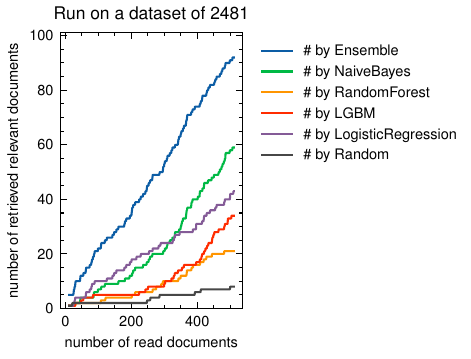}

}

\caption{\label{fig-example-run}An example run for 500 iterations on a
dataset. The \emph{Ensemble} curve shows the number of documents that
have been found by the overall system. The other curves display the
number of relevant documents that have been found by the individual
members within \(\mathcal{C}\). The reader may notice that curves start
slightly after 0 documents and end slightly after 500 documents. This is
caused by the fact that our method requires five relevant and five
irrelevant documents at the start of the process (see
Section~\ref{sec-training-ranking-sampling}), which results in this
shift.}

\end{figure}%

\begin{table}

\caption{\label{tbl-fstats}Frequency statistics for the example run in
Figure~\ref{fig-example-run}.}

\centering{

\begin{tabular}{lrrrrrr}
\toprule
$f_0$ & $f_1$ & $f_2$ & $f_3$ & $f_4$ & $f_5$ & $n$\\
\midrule
? & 40 & 33 & 17 & 2 & 0 & 92\\
\bottomrule
\end{tabular}

}

\end{table}%

We execute the procedure described in
Section~\ref{sec-pse-without-multiple-reviewers} on the dataset
collected for a systematic review {[}29{]}. This dataset is part of the
test collection of {[}37{]}. This dataset contains 2481 documents, of
which 120 are relevant, so the ground truth for \(N = 120\). We simulate
500 iterations (i.e., 500 review decisions) using our methodology. At
iteration \(t=500\), we have the following frequency statistics,
displayed in Table~\ref{tbl-fstats}. We omit matrix \(\mathbf{X}\) for
practical reasons.

At iteration \(t = 500\), the number of retrieved relevant documents
\(n = 92\). Using the frequency statistics, we can use Chao's moment
estimator to obtain a point estimate of the total number of omitted
relevant documents \(\hat{f}_0\) and the resulting total number of
relevant documents \(\hat{N}\). Chao's estimator is formulated as
follows, \begin{equation}\phantomsection\label{eq-chao}{
\hat{N} = n + \hat{f}_0,\quad \hat{f}_0 = \left\{ \begin{array}{llr}
  \frac{f_1^2}{2f_2} & \textrm{if} & f_2 > 0 \\
  \frac{f_1\left(f_1 - 1\right)}{2\left(f_2 +1\right)} & \textrm{if} & f_2 =0
\end{array} \right. \ .
}\end{equation} For the derivation of Equation~\ref{eq-chao} for the
case \(f_2>0\), we refer to {[}10 - 786{]}. The case for \(f_2 =0\) is
needed as the upper formula cannot be calculated when \(f_2 = 0\) due to
a division by zero. In {[}11{]}, an adjusted formula is given for when
\(f_2=0\). For the example data, the estimate is
\[\hat{N} = 92 + \frac{40^2}{2 \cdot 33} = 116.24\ ,\] which differs
3.76 from the true value of \(N = 120\).

\subsubsection{Confidence Interval}\label{confidence-interval}

In order to increase the reliability of the stopping criterion, it would
be beneficial to have a measure of confidence for our point estimates.
Stopping at a moment when the variance is high is not ideal. As in
{[}31{]}, we introduce a \emph{conservative} stopping criterion, which
uses the upper bound of a 95 \% confidence interval as the stopping
criterion (we describe the criteria in more detail in
Section~\ref{sec-stopping-criterion}). Chao {[}10, 11{]} provides the
following variance estimator.
\begin{equation}\phantomsection\label{eq-chao-variance}{
\hat{\sigma}^{2}_{\hat{N}} = \left\{ \begin{array}{llr}
f_2 \left(\frac{1}{4} \left(\frac{f_1}{f_2}\right)^4 + \left(\frac{f_1}{f_2}\right)^3 + \frac{1}{2} \left(\frac{f_1}{f_2}\right)^2 \right) & \textrm{if} & f_2 > 0 \\
\frac{{f_1}\left({f_1}-1 \right)}{2} + \frac{{f_1}\left(2 f_1 -1 \right)^2}{4} - \frac{f_1^4}{4\hat{N}}&\textrm{if} & f_2 = 0
\end{array} \right. \ .
}\end{equation} \noindent Then the confidence interval can be estimated
as, \begin{equation}\phantomsection\label{eq-chao-ci1}{
\left[n + \frac{\hat{N}-n}{Q}, n +  \left(\hat{N} -n\right)Q\right]\ ,
}\end{equation} where
\begin{equation}\phantomsection\label{eq-chao-ci2}{
Q = e^{1.96\sqrt{\ln{\left(1 + \frac{\hat{\sigma}^2_{\hat{N}}}{\left(\hat{N}-n\right)^2}\right)}}}\ ,
}\end{equation} in which is 1.96 is the critical value of the normal
distribution. For the case of \(f_2 > 0\) in
Equation~\ref{eq-chao-variance}, we refer to {[}10{]}, for the case of
\(f_2 = 0\), we refer to {[}11{]}\footnote{In {[}11{]}, the equation for
  variance for both cases contains a term \(k\). Conform {[}11, Equation
  (6a){]} we rewrite the formula for both cases to a version without
  this term \(k\) so that it matches the equation for variance in
  {[}10{]}.}.

\noindent For the example data, the variance is \(100.82\). Then,
according to Equation~\ref{eq-chao-ci1} and Equation~\ref{eq-chao-ci2},
the 95 \% CI for this data is \(\left[103.11, 144.88 \right]\). In
section Section~\ref{sec-stopping-criterion}, we further detail how the
point estimates and intervals are used to stop the review process.

\subsubsection{Model assumptions}\label{model-assumptions}

Chao's estimator is based on the \emph{heterogeneity model} \(M_h\)
introduced in {[}6, 7{]}. The model \(M_h\) assumes that the capture
probability only varies among the individuals (so, the relevant
documents in our case). Some documents have a higher probability of
being selected by the committee members than others. Chao {[}10{]}
assumes that for (\(p_{ij} = p_i\)), where \(i = 1, 2, \ldots, N\) and
\(j = 1, 2, \ldots, C\), then, \(p_1, p_2, \ldots, p_N\) are a random
sample from an unknown probability distribution function. Chao's
estimator assumes that the number of observations of an individual (in
our case, the number of committee members that have found a relevant
document \(i\)) is a realization from a zero-truncated Poisson
distribution. Therefore, it is crucial to verify that the number of
observations of a relevant document (which can be one of
\(\left\{1,2,3,4,5 \right\}\)) can be assumed to be a realization of a
(truncated) Poisson distribution.

Poisson originally formulated his distribution as a limit of the
binomial distribution {[}23{]} with a success probability \(p\) and
\(N\) realizations, with \(N\) approaching infinity, \(p\) tending to
zero, and \(Np\) remaining finite and equivalent to the Poisson
parameter \(\lambda\). However, even when \(N\) is small, the Poisson
distribution can reasonably approximate the binomial distribution, given
that \(p\) is small enough {[}38{]}. In our case, \(N\) is small, and
the chance of encountering a document is also small.

The Poisson parameter can vary for each document \(i\), allowing for
heterogeneity in capture probabilities. This is convenient as some
documents may be harder to find than others. Moreover, some methods may
be better suited to finding a particular document than others. For
example, the predictions of the Logistic Regression classifier may
differ from the predictions of a Random Forest, even when the same set
of documents are given as training data. In this case, for a document
\(i\) the number of observations is stated as being
\(\lambda_i = \lambda_{i,\textrm{LR}} + \ldots + \lambda_{i,{\textrm{RF}}}\).
\(\lambda_i\) is a Poisson parameter \(\lambda_{i,\textrm{LR}}\) and
\(\lambda_{i,{\textrm{RF}}}\) are also Poisson parameters. This follows
from a property of the Poisson distribution found by {[}12{]}; a Poisson
is infinitely divisible. If you have two independent Poisson random
variables, \(X_1\) with parameter \(\lambda_1\) and \(X_2\) with
parameter \(\lambda_2\), then the sum of these two random variables,
\(X_1 + X_2\), will also follow a Poisson distribution with parameter
\(\lambda_1 + \lambda_2\). In our case, each relevant document \(i\) has
Poisson parameter \(\lambda_{ij}\) where \(j\) is one of the members of
the committee which uses a specific classification algorithm. The sum
of, in our case, the five Poisson parameters leads to a Poisson
parameter for each document \(i\):
\(\lambda_i = \sum_{j \in C} \lambda_{ij}\).

An issue that sometimes arises in Population Size Estimation and
Capture-recapture studies is \emph{contagion}, which happens when the
capture of an individual changes the probability of capturing it a
second time (for example, an animal may change its behavior after
capture, or the researcher may become better at finding that specific
animal after observing it). Contagion violates the Poisson assumption
(which follows from the independence between the trials in the binomial
distribution). However, as our committee members search independently,
we know that the contagion problem is not present: the Poisson
parameters \(\lambda_{ij}\) for other members \(C_j\) do not change
after its capture by any other member.

Given the properties of our problem and the Poisson distribution, we can
state that this distribution is suitable and fulfills the assumptions of
Chao's estimator. Accounting for the heterogeneity, Chao's estimator
provides a lower bound on the number of relevant documents. However, a
simulation study in {[}10{]} showed that, in many cases, it is a good
estimator for \(N\) in general. Notice that Chao's estimator only uses
the frequencies of the documents discovered once and twice. The
intuition behind the Chao estimator is that (for the ecology use-case)
if you have seen many animals once (relative to the number of animals
seen twice), then probably there are a lot more that you have missed
completely; it would be surprising if you would have seen all unique
animals exactly once. The more animals you have seen twice (relative to
those seen once), the larger the probability you have seen most of them.
Chao's estimator formalizes this intuition and provides a lower bound by
only considering the number of individuals seen once and twice.

\subsection{Poisson Regression
version}\label{poisson-regression-version}

As mentioned earlier, we also use a Poisson regression version of Chao's
Moment estimator as presented by Rivest et al. {[}32{]}. This model also
takes the frequencies \(f_{3,4,5}\) into account, in addition to the
frequencies \(f_{1,2}\). Moreover, this model enables us to obtain a 95
\% confidence interval using the profile likelihood instead of the
asymptotic approach from {[}10{]}. The profile likelihood method has
been advocated by many statisticians {[}1, 17, 19, 21{]}. We describe
this method below. We will refer to this method as \emph{Chao (Rivest)},
while we will refer to Chao's Moment Estimator as \emph{Chao (1987)}.

Using the data in Table~\ref{tbl-fstats}, we can specify the design
matrix for the \emph{Chao (Rivest)} model in Table~\ref{tbl-designmat}.
The model has \(C-2\) parameters, called \(\eta\) parameters, for
modeling heterogeneity in capture probabilities within the set of
relevant documents. In our case, as \(C = 5\), we have 3 \(\eta\)
parameters. The \(Y\) variable contains the frequency statistics (from
top to bottom, \(f_5\) to \(f_1\)).

\begin{table}

\caption{\label{tbl-designmat}The data \((Y)\), which contains \(f_5\)
to \(f_1\) (above to below). The rest of the columns belong to the
design matrix for the \emph{Chao (Rivest)} model.}

\centering{

\centering
\begin{tabular}[t]{rrrrrr}
\toprule
\textbf{$Y$} & \textbf{Intercept $(\gamma)$} & \textbf{beta $(\beta)$} & \textbf{eta3 $(\eta_3)$} & \textbf{eta4 $(\eta_4)$} & \textbf{eta5 $(\eta_5)$}\\
\midrule
\cellcolor{gray!15}{0} & \cellcolor{gray!15}{1} & \cellcolor{gray!15}{5} & \cellcolor{gray!15}{3} & \cellcolor{gray!15}{2} & \cellcolor{gray!15}{1}\\
2 & 1 & 4 & 2 & 1 & 0\\
\cellcolor{gray!15}{17} & \cellcolor{gray!15}{1} & \cellcolor{gray!15}{3} & \cellcolor{gray!15}{1} & \cellcolor{gray!15}{0} & \cellcolor{gray!15}{0}\\
33 & 1 & 2 & 0 & 0 & 0\\
\cellcolor{gray!15}{40} & \cellcolor{gray!15}{1} & \cellcolor{gray!15}{1} & \cellcolor{gray!15}{0} & \cellcolor{gray!15}{0} & \cellcolor{gray!15}{0}\\
\bottomrule
\end{tabular}

}

\end{table}%

We use the package RCapture {[}32{]} to fit the model, which uses a
standard Generalized Linear Model fitting algorithm. Given the data in
Table~\ref{tbl-fstats}, this algorithm fits the model with the following
parameters as presented in the first half (``Before removal'') of
Table~\ref{tbl-fittedcoefs-etaremoved}. In this algorithm, all \(\eta\)
parameters fitted with a negative coefficient are set to zero, as these
parameters should theoretically be greater than or equal to zero
{[}32{]} (when set to zero, these parameters are effectively removed
from the design matrix). Note that after setting an \(\eta\) parameter
to zero and fitting a new model, the other \(\eta\) parameters could be
fitted with a negative coefficient, so this process is repeated until
all \(\eta\) parameters are positive or removed. For this data, the
algorithm does indeed remove all \(\eta\) parameters. The parameters of
the final model are presented in the second half (``After removal'') of
Table~\ref{tbl-fittedcoefs-etaremoved}.

\begin{table}

\caption{\label{tbl-fittedcoefs-etaremoved}The fitted coefficients from
the data presented in Table~\ref{tbl-designmat} before and after removal
of negative \(\eta\) parameters.}

\centering{

\centering
\begin{tabular}[t]{lrrrrrr}
\toprule
\multicolumn{1}{c}{ } & \multicolumn{3}{c}{Before removal} & \multicolumn{3}{c}{After removal} \\
\cmidrule(l{3pt}r{3pt}){2-4} \cmidrule(l{3pt}r{3pt}){5-7}
\textbf{ } & \textbf{Est.} & \textbf{S.E.} & \textbf{p} & \textbf{Est.} & \textbf{S.E.} & \textbf{p}\\
\midrule
\cellcolor{gray!15}{Intercept $(\gamma)$} & \cellcolor{gray!15}{\num{3.19}} & \cellcolor{gray!15}{\num{0.36}} & \cellcolor{gray!15}{\num{<0.001}} & \cellcolor{gray!15}{\num{3.50}} & \cellcolor{gray!15}{\num{0.22}} & \cellcolor{gray!15}{\num{<0.001}}\\
beta $(\beta)$ & \num{0.50} & \num{0.24} & \num{0.033} & \num{0.29} & \num{0.11} & \num{0.010}\\
\cellcolor{gray!15}{eta3 $(\eta_3)$} & \cellcolor{gray!15}{\num{-0.07}} & \cellcolor{gray!15}{\num{0.45}} & \cellcolor{gray!15}{\num{0.885}} & \cellcolor{gray!15}{} & \cellcolor{gray!15}{} & \cellcolor{gray!15}{}\\
eta4 $(\eta_4)$ & \num{-1.19} & \num{0.87} & \num{0.174} &  &  & \\
\cellcolor{gray!15}{eta5 $(\eta_5)$} & \cellcolor{gray!15}{\num{-20.63}} & \cellcolor{gray!15}{\num{42247.17}} & \cellcolor{gray!15}{\num{1.000}} & \cellcolor{gray!15}{} & \cellcolor{gray!15}{} & \cellcolor{gray!15}{}\\
\bottomrule
\end{tabular}

}

\end{table}%

Using the parameters in Table~\ref{tbl-fittedcoefs-etaremoved}, we can
calculate the estimate for the number of relevant documents as follows;

\[\hat{N} = n +e^{\hat{\gamma}} = 92 + e^{3.5} = 125.18\  .\]This value
differs differs by \(5.18\) from the ground truth \(N = 120\) and is
higher than the estimate by \emph{Chao (1987)} (\(116.24\)).

\subsubsection{Confidence Interval}\label{sec-methodology-ci-poisson}

The confidence interval for \emph{Chao (Rivest)} is calculated by using
the deviance or log-likelihood ratio of the models from complete tables,
introduced in {[}17{]}. This procedure is set up as follows. Suppose we
have an incomplete table with only the observed counts (for example,
Table~\ref{tbl-designmat}). We can extend and complete this table by
adding a row for the unobserved count \(u\). Then, we need to find the
Poisson model for the extended table \(PE\) with the lowest deviance. We
can find this by a search for \(u\) in the interval
\(u \in\left[0, \frac{3}{2}\hat{f_0} \right]\) , equivalently
\(\hat{N}_C = n + u\), for a conditional estimate for \(N\) based on the
complete table. We record for each model the log likelihood for

\[
L(\hat{N}_C, \hat{\theta}_{N_C}; n) = D_{PE} - 2 {ct},
\] where \(D_{PE}\) is the deviance for model \(PE\) and a correction
term {[}17, 32{]}. \[
ct = \left\{
  \begin{array}{lr}
        u - \hat{N}_C - \frac{\log{\frac{u}{\hat{N}_C}}}{2}, & \text{if } \hat{N}_C > 100 \wedge u \geq 2 \\
        -\hat{N}_C + \frac{\log{2\pi \hat{N}_C}}{2},        & \text{if } \hat{N}_C > 100 \wedge u \in [0,1] \\\
        \log{\frac{u^u \cdot \hat{N}_C!}{\hat{N}_C^{\hat{N}_C}\cdot u!}},       & \text{otherwise}
  \end{array}
  \right. \  .
\]Then, we find the value \(\hat{u}^\star\) that maximizes this
log-likelihood (or minimizes the deviance). By using the asymptotic
\(\chi^2_1\) distribution, we can find the values \(u\) that increase
this value by an amount \(k_\alpha\), where \(k_\alpha\) is a critical
value calculated using the quantile function from this \(\chi^2_1\)
distribution {[}32{]}. In this case, for a 95 \% CI, the critical value
\(k_{\alpha = 0.05} = 3.84\).

\begin{figure}

\centering{

\includegraphics{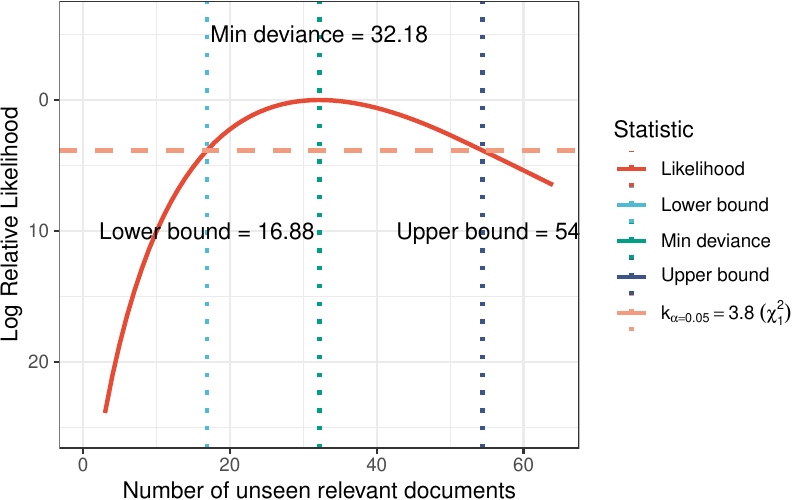}

}

\caption{\label{fig-loglikehoodprofile}Calculating the 95 \% confidence
interval using the profile likelihood method for the frequency
statistics in Table~\ref{tbl-fstats}. In this figure, the log-likelihood
for \(u^\star\) is subtracted from the likelihood in aid of the
visualization. Additionally, we inverted the \(y\)-axis for this
purpose. By finding the values \(u\) that intersect with the line
\(y = k_{\alpha=0.05} = 3.84\), we can find the lower bound and upper
bound of the interval.}

\end{figure}%

This procedure is visualized in Figure~\ref{fig-loglikehoodprofile} for
the data in Table~\ref{tbl-fstats}. Combining the interval found for
\(\hat{f_0}\) with \(n = 92\), the 95 \% CI obtained using this method
is \([108.88, 146.39]\), which is similar to the interval obtained using
\emph{Chao (1987)} (\(\left[103.11, 144.88 \right]\)).

\subsection{Stopping Criterion}\label{sec-stopping-criterion}

Using the estimate \(\hat{N}\) and the corresponding 95 \% CI for \(N\),
we can determine if we can terminate the TAR procedure. The user can
specify a recall target, such as 95 \% recall (note that the 95\% recall
target is not be confused with the 95 \% of the CI). The system tracks
the estimate and CI to determine if the stopping criterion has been met.
However, there are multiple ways to decide on the recall statistics and
estimates.

In our implementation, similar to {[}31{]}, we use the estimates as
described in the previous sections in the following two ways:

\begin{description}
\item[Conservative.]
We use the \emph{upper bound} of the CI \(\hat{N}_\textrm{sup}\) to
determine the current recall estimate. The current recall estimate of
iteration \(t\) is defined as
\(\hat{R}_t = \frac{|\mathcal{L}^+_t|}{\hat{N}_{\textrm{sup}}}\).
\item[Optimistic.]
Here, we use the \emph{point estimate} \(\hat{N}\) of the estimator as
to determine the current recall estimate. For this criterion, the
current recall of iteration \(t\) estimate is defined as
\(\hat{R}_t = \frac{|\mathcal{L}^+_t|}{\hat{N}}\).
\end{description}

Both methods are triggered when
\(\hat{R}_{t} \geq R_{\textrm{target}}\). The estimated recall
percentage is rounded to nearest integer value to allow some numerical
imprecision. Moreover, the criteria can also only be triggered after
\(|\mathcal{L}_t| > 100\), as the estimates may fluctuate heavily in the
first phase of the procedure.

Combined with the two estimators, we provide four stopping criteria,
which can be used with a user specified recall target.

\begin{itemize}
\item
  \emph{Chao (1987)} - Conservative
\item
  \emph{Chao (1987)} - Optimistic
\item
  \emph{Chao (Rivest)} - Conservative
\item
  \emph{Chao (Rivest)} - Optimistic
\end{itemize}

\subsection{Active Learning procedure}\label{active-learning-procedure}

In the following sections, we describe the Machine Learning and Active
Learning aspects of our method. First, we briefly describe the feature
extraction method that we employ, followed by the classifiers that are
used within the ensemble. As the data is often imbalanced, dynamic
resampling is used as a balancing procedure. Finally, we describe the
query strategy and batching scheme.

\subsubsection{Feature Extraction}\label{sec-methodology-fe}

All documents are represented as TF-IDF vectors for all classification
algorithms. We only include terms with a minimum document frequency of 2
and limit the term matrix to 3000 terms by selecting based on the term
frequency across the dataset in question. Moreover, English stop words
are excluded from the term matrix. If desired, our framework allows
substituting the TF-IDF vectorizer with another algorithm; for example,
Doc2Vec or SentenceBERT on the committee member level, allowing the use
of multiple vector representations concurrently. However, in our
experiments, we will not use this feature.

\subsubsection{Classifiers}\label{classifiers}

We create an ensemble of various learners, each of which uses a unique
classification algorithm. The decision boundaries of each algorithm will
differ, resulting in a different order of document selection. The
algorithms we use are:

\begin{description}
\item[Multinomial Naive Bayes.]
Multinomial Naive Bayes is a probabilistic classification algorithm
frequently used for text classification tasks. It is also used in TAR
systems, e.g., it is the default classification algorithm in {[}37{]}.
In most cases, this algorithm is used with documents in Bag-of-Words
representation, such as TF-IDF vectors.
\item[Logistic Regression.]
Logistic Regression is a common classification algorithm used in TAR,
for instance, in AutoTAR {[}15{]} and derivatives. The method is also
used or available as an option in {[}31, 37{]}.
\item[Random Forest.]
Random Forest {[}22{]} is an ensemble learning method that uses multiple
decision trees to make predictions. Each tree in the forest is trained
on a randomly selected bootstrap sample of the training data, and at
each node, the best split is chosen among a randomly selected subset of
the features. The final prediction is made by a majority vote among the
trees in the forest. Ranking is possible by using the mean predicted
class probabilities of the trees in the forest. The class probability of
a single tree is defined as the fraction of training samples in the leaf
that have the same class as the leaf. The Random Forest classifier is
available as an option in {[}37{]}.
\item[Light Gradient Boosting Machine (LGBM)]
Light GBM {[}28{]} is a gradient boosting framework that uses a
tree-based learning algorithm. It is designed to be highly efficient and
scalable, with faster training speeds and lower memory usage compared to
other popular gradient boosting frameworks. To our knowledge, it is not
used in any existing TAR systems, but it performs similarly to the
Random Forest method for some datasets.
\item[Random Sampling]
We also include one member that does not use machine learning nor active
learning. The idea behind this is that we may capture instances in
unexplored areas of the search space that are not covered by the greedy
searching machine learning-based committee members.
\end{description}

Support Vector Machines (SVM) is another viable option to consider
instead of one of the previously mentioned machine learning methods.
However, during our initial experiments, we discovered that using SVM
significantly impacted the training time for each iteration and, thus,
the total runtime of each experiment. In practical applications, this
may not be a concern, as manual review time typically exceeds the
training time of the models. We anticipate substituting one of the
classifier algorithms with SVM will not substantially impact the
results.

\subsubsection{Balancing}\label{balancing}

Many classification algorithms encounter difficulties when fitting
models with limited data, especially in the case of imbalanced datasets.
This limitation may be because the prevalence of the relevant class is
generally low in most TAR datasets. To address this challenge, one
potential solution is to balance the training data. One method called
dynamic resampling {[}20, 37{]}, rebalances the training data by
oversampling documents from the positive class \(\mathcal{L^+}\) and
undersampling from \(\mathcal{L^-}\). The amount of oversampling and
undersampling is dynamic and depends on the sizes of the sets
\(\mathcal{L}^-, \mathcal{L}^+\) and \(\mathcal{L}
\). The methods ensures that the size of the training data remains the
same in terms of \(|\mathcal{L}|\). A more detailed description of this
method is given in {[}20{]}.

Our early experiments have demonstrated that using this method
significantly improves the performance of our models, particularly in
terms of WSS@95 (Work Saved over Sampling at 95 \% recall), leading us
to employ this procedure to balance the training data for all the
classifier models in our system.

\subsubsection{Training, Ranking, and
Sampling}\label{sec-training-ranking-sampling}

Each of the members' models is trained using their own labeled sets
\(\mathcal{L}_i\), and the training data is balanced using the method
described previously. Then, we predict for each document within the
member's \(\mathcal{U}_i\) its relevancy probability. These scores are
used to rank each document. Each member \(i\) prepares a batch of
\(b_i\) documents, which are greedily sampled from a batch which
consists of the top-\(b_i\) documents from the ranking produced by the
model. Then, when the user queries a document from the committee, one of
the members is randomly selected to propose a document, which will be
the first document in the batch. When all documents from a member's
batch have been labeled, its model is retrained. After retraining, we
update the batch size \(b_i\) in the same manner as in AutoTAR {[}15{]}:
\(b_i \gets b_i + \lceil \frac{b_i}{10}\rceil\). The initial batch size
\(b_i = 1\).

This process is repeated until the stopping criterion is met. Each
committee member needs one document from each class as initial training
data to start the process. As there are five members in our committee,
we need five relevant documents in total (and five non-relevant ones).

\bookmarksetup{startatroot}

\section{Experimental Setup}\label{sec-exp-setup}

Below, we describe our experimental setup and research questions. We
also briefly describe the datasets that are used for benchmarking.
Furthermore, we list the existing methods that we take into account in
our comparison.

\subsection{Research questions}\label{sec-research-questions}

In TAR, the goal is to retrieve as many relevant documents as possible
(achieving a high recall) while minimizing the workload in terms of
review work. A good stopping criterion should achieve its recall target
with minimal cost. In order to evaluate how our method performs (as well
as compared to other methods), we study the following research
questions.

\begin{enumerate}
\def\labelenumi{\arabic{enumi}.}
\tightlist
\item
  How does our Active Learning strategy perform in terms of WSS@95 and
  WSS@100 compared to other methods?
\item
  Can our stopping criteria help the user achieve its recall target in a
  timely fashion?
\item
  How reliable are our stopping criteria?
\item
  How do our stopping criteria compare to other methods that estimate
  the number of relevant documents?
\item
  How do our stopping criteria compare to other methods that do not
  provide such an estimate?
\end{enumerate}

\subsection{Study design}\label{sec-study-design}

To ensure that our findings are generalizable to new and unseen
datasets, we run each Active Learning method and corresponding stopping
criteria on a large collection of datasets of various domains. Moreover,
we will repeat the experiment multiple times for each datasets. The
datasets are described in more detail in
Section~\ref{sec-datasets-overview}.

\subsubsection{Active Learning
initialization}\label{active-learning-initialization}

Many TAR procedures require a seed set of relevant and irrelevant
documents to start the Active Learning loop. Earlier work has shown that
the documents in the seed set can influence the results {[}8{]}. Our
method, consisting of several committee members (each representing an
active learning strategy), requires a seed set of five relevant and five
irrelevant documents. To ensure that our results do not depend on a
single seed set, we repeat our experiments with varying sets of seed
documents. We use 30 distinct seed sets for each dataset and method. To
ensure a fair comparison, the methods that can work with a smaller seed
set also get a seed set of the same size as our method. Moreover, the
set of sampled documents depends on the seed value that is given to the
Pseudo Random Generator; this means that an experiment for a method
\(A\) with seed \(s\) and an experiment with method \(B\) with the same
seed value \(s\) use the same documents to initialize the Active
Learning procedure.

\subsubsection{Feature Extraction}\label{feature-extraction}

All methods in our study use TF-IDF feature vectors. To ensure a fair
comparison, we keep the configuration the same for each method, so as
per Section~\ref{sec-methodology-fe}, limited to 3000 terms after
filtering English stop words from the vocabulary.

\subsubsection{Evaluation metrics}\label{evaluation-metrics}

We let each algorithm run until all documents are screened. During the
experiment, each criterion can signal when it is triggered. Moreover, if
the method produces an estimate for the size of \(\mathcal{D}^+\), then
this estimate is also registered. When a method triggers a stopping
criterion, the following metrics are recorded.

\begin{description}
\item[Effort.]
The percentage of documents that have been screened after triggering the
stopping criterion.

\begin{equation}\phantomsection\label{eq-effort}{
E = \frac{|\mathcal{L}|}{|\mathcal{D}|}
}\end{equation}
\item[Recall.]
The percentage of relevant documents that have been found based on the
\emph{a priori} knowledge from the ground truth dataset. We will record
the recall when the stopping criterion has been satisfied.
\begin{equation}\phantomsection\label{eq-recall}{R =\frac{
        |\mathcal{L}^{+}|
      }{
        |\mathcal{L}^{+} \cup \mathcal{U}^{+}|
      }}\end{equation}
\item[Recall Error.]
For the methods that can specify a recall target; the error is the
absolute difference between the achieved recall and the target recall
when the stopping criterion is triggered, divided by the recall target.
\begin{equation}\phantomsection\label{eq-recallerror}{RE =\frac{
        |R_{\textrm{stop}} - R_{\textrm{target}}|
      }{R_{\textrm{target}}}}\end{equation}
\item[Work Saved over Sampling.]
This metric expresses the work reduction over random sampling. We
calculate this as follows:
\begin{equation}\phantomsection\label{eq-wss}{
      WSS =
      \frac{
        |\mathcal{U}|
      }{
        |\mathcal{D}|
      } - \left(
      1 -
      \frac{
        |\mathcal{L}^{+}|
      }{
        |\mathcal{L}^{+} \cup \mathcal{U}^{+}|
      }\right)}\end{equation}
\item[loss\(_\textrm{er}\).]
This metric introduced in {[}15{]} aims to assess both review costs and
recall. It is defined as:

\begin{equation}\phantomsection\label{eq-loss-er}{
\textrm{loss}_\textrm{er} = \left(1 - R\right)^2 + {\left(\frac{100}{|\mathcal{D}|} \right)}^2 \cdot {\left(\frac{|\mathcal{L}|}{|\mathcal{L}^+| + 100}\right)}^2\quad,
}\end{equation}

where \(R\) is the recall as defined in Equation~\ref{eq-recall}. This
metric consists of two term. The first is the loss due to missing
relevant documents, which becomes higher when the recall is low. The
second term is the loss in terms of effort. The two scalar values 100 in
the metric are considered a correction for reasonable extra work for
achieving a high recall.
\item[Target met.]
In this metric, we test for each individual run if the recall target was
met. We report the percentage of runs over all datasets in which this is
the case.
\item[Triggered.]
For each individual run, we record if the stopping criterion was
triggered before all documents were exhausted. As with the \emph{Target
met} metric, we report the percentage of runs over all datasets in which
this is indeed the case.
\end{description}

Running the experiment till exhaustion enables us to measure what would
have happened if the method did not stop. This enables us to assess how
many documents would have remained for a method to achieve its target
recall in case it stopped the process too early.

\subsection{Datasets}\label{sec-datasets-overview}

We use benchmark datasets of several corpora to ensure the results of
our methods are generalizable to unseen datasets. One of the corpora
used in our experiments consists of systematic literature reviews from
{[}18{]}. This corpus contains datasets with inclusion and exclusion
records from several real-world published systematic reviews from
various domains: psychology, the medical field, and information
sciences, among others. We also include in our experiments three corpora
from the Conference and Labs of the Evaluation Forum (CLEF)
Technology-Assisted Reviews in Empirical Medicine datasets from the
years 2017, 2018, and 2019 {[}24--26{]}. This corpus is in TREC format.
The CLEF Task was aimed to evaluate search methods aimed to identify all
relevant works for a systematic literature review in empirical medicine.

Table~\ref{tbl-datasets-overview} shows some of the dataset
characteristics of each of the corpora. We describe the individual
datasets in more detail in Appendix~\ref{sec-dataset-statistics-for-slr}
in Table~\ref{tbl-datasets-slr}, Table~\ref{tbl-datasets-clef2017},
Table~\ref{tbl-datasets-clef2018}, and
Table~\ref{tbl-datasets-clef2019}.

\begin{table}

\caption{\label{tbl-datasets-overview}Main statistics of the corpora
included in our experiments. Here, \(N\) is the number of
datasets/topics within each corpus. We report the median and the
interquartile range of the dataset statistics within each corpus.}

\centering{

\centering
\begin{tabular}{lcccc}
\toprule
 & \textbf{SYNERGY}, N = 20 & \textbf{clef2017}, N = 31 & \textbf{clef2018}, N = 24 & \textbf{clef2019}, N = 19\\
\midrule
\cellcolor{gray!15}{\textbf{\# Relevant}} & \cellcolor{gray!15}{67 (32, 106)} & \cellcolor{gray!15}{92 (49, 126)} & \cellcolor{gray!15}{67 (39, 277)} & \cellcolor{gray!15}{64 (33, 78)}\\
\textbf{\# Irrelevant} & 3,554 (1,690, 6,864) & 3,211 (1,498, 6,950) & 5,064 (1,700, 8,553) & 3,158 (1,770, 5,412)\\
\cellcolor{gray!15}{\textbf{Size}} & \cellcolor{gray!15}{3,577 (1,725, 7,329)} & \cellcolor{gray!15}{3,241 (1,596, 7,261)} & \cellcolor{gray!15}{5,123 (1,898, 8,592)} & \cellcolor{gray!15}{3,169 (1,840, 5,506)}\\
\textbf{Prevalence (\%)} & 1.5 (0.8, 5.0) & 2.4 (1.0, 5.6) & 2.8 (1.0, 6.6) & 1.8 (0.9, 5.3)\\
\bottomrule
\end{tabular}

}

\end{table}%

\subsubsection{Selection}\label{selection}

As said earlier, our method needs at least five initial relevant
documents to train a machine learning model for each committee member.
Moreover, methods such as the Target rely on the fact that there are at
least ten documents within the dataset. Therefore, we opt to only
include datasets in our experiments that contain at least ten relevant
documents because the inclusion of runs in which more than half of all
relevant documents as prior knowledge may distort the results. The CLEF
corpora contain datasets with an extremely small size. We decided to
exclude datasets with less than 500 documents for two reasons: first,
many criteria do not work well on these datasets (e.g., Stop200,
Stop400, Budget, Knee, Conservative, and Optimistic). The second reason
is that the advantage of using TAR on these datasets is minimal compared
to the work required to process the whole dataset. Furthermore, we
exclude two datasets from the CLEF corpora due to their large size; we
cannot assess the standard version of AUTOSTOP on these datasets due to
memory limitations (see Section~\ref{sec-rl-autostop}).

\subsection{Comparison to other
methods}\label{comparison-to-other-methods}

To assess our method, we will compare the metrics to methods presented
in earlier work. We will include several methods discussed in
Section~\ref{sec-stopping-criteria-relatedwork}. An overview of all the
methods included in our experiments is given in
Table~\ref{tbl-exp-methods}.

\begin{longtable}[]{@{}
  >{\raggedleft\arraybackslash}p{(\columnwidth - 8\tabcolsep) * \real{0.2000}}
  >{\centering\arraybackslash}p{(\columnwidth - 8\tabcolsep) * \real{0.2000}}
  >{\centering\arraybackslash}p{(\columnwidth - 8\tabcolsep) * \real{0.2000}}
  >{\centering\arraybackslash}p{(\columnwidth - 8\tabcolsep) * \real{0.2000}}
  >{\raggedleft\arraybackslash}p{(\columnwidth - 8\tabcolsep) * \real{0.2000}}@{}}
\caption{An overview of all methods included in the experiments. The
\emph{I}, \emph{S}, and \emph{H} in the \emph{Applicability} column
stand for \emph{Interventional, Standoff,} and \emph{Hybrid}, whereas
the \emph{C} and \emph{H} in the \emph{Certification} column stand for
\emph{Certification} and \emph{Heuristic}. The \emph{AL} column gives
the \emph{Active Learning} method that is used with the criteria in the
experiments. The CMH methods are the Hypergeometric methods from {[}8{]}
referring to the first letters of the authors' surnames. Our
implementation of some methods is based on the implementation in TARexp
{[}42{]}, if this is the case it is
listed.}\label{tbl-exp-methods}\tabularnewline
\toprule\noalign{}
\begin{minipage}[b]{\linewidth}\raggedleft
\textbf{Method}
\end{minipage} & \begin{minipage}[b]{\linewidth}\centering
\textbf{Applicability}
\end{minipage} & \begin{minipage}[b]{\linewidth}\centering
\textbf{Certifcation}
\end{minipage} & \begin{minipage}[b]{\linewidth}\centering
\textbf{AL}
\end{minipage} & \begin{minipage}[b]{\linewidth}\raggedleft
\textbf{Source}
\end{minipage} \\
\midrule\noalign{}
\endfirsthead
\toprule\noalign{}
\begin{minipage}[b]{\linewidth}\raggedleft
\textbf{Method}
\end{minipage} & \begin{minipage}[b]{\linewidth}\centering
\textbf{Applicability}
\end{minipage} & \begin{minipage}[b]{\linewidth}\centering
\textbf{Certifcation}
\end{minipage} & \begin{minipage}[b]{\linewidth}\centering
\textbf{AL}
\end{minipage} & \begin{minipage}[b]{\linewidth}\raggedleft
\textbf{Source}
\end{minipage} \\
\midrule\noalign{}
\endhead
\bottomrule\noalign{}
\endlastfoot
\textbf{AutoStop} & I & C & AutoStop & {[}31{]} \\
\textbf{Budget} & S & H & AutoTAR & {[}15, 42{]} \\
\textbf{Chao (ours)} & I & C & Ensemble & \\
\textbf{CMH-Standoff} & S & C & AutoTAR & {[}8, 42{]} \\
\textbf{CMH-Hybrid} & H & C & AutoTAR & {[}8{]} \\
\textbf{Half} & S & H & AutoTAR & {[}42{]} \\
\textbf{Knee} & S & H & AutoTAR & {[}15, 42{]} \\
\textbf{Quant (CI)} & S & C & AutoTAR & {[}41, 42{]} \\
\textbf{Rule2399} & S & H & AutoTAR & {[}42{]} \\
\textbf{Stop after} \(k\) & S & H & AutoTAR & \\
\textbf{Target} & H & H & AutoTAR & {[}15{]} \\
\end{longtable}

For the standoff methods, we choose to apply them to AutoTAR, as this
method is considered state of the art and does not perform any
additional work to decide when to stop. For AutoStop, we specifically
implemented the Horvitz-Thompson variant, as suggested by the authors
{[}31{]}.

\subsection{Implementation}\label{implementation}

We provide a Python library, \texttt{python-allib} (see {[}5{]}), which
implements our methods and all the baselines, some of which are adapted
from TARexp {[}42{]}. The TARexp package only allows the comparison of
stop criteria that fall in the standoff category (see
Section~\ref{sec-stopping-criteria-relatedwork}), but not methods of
interventional nature. Our framework allows various forms of ranking and
arranging the reviewer workload so interventional methods can be
implemented. The library is based on the Python package
\texttt{instancelib} {[}4{]}, enabling integration within annotation
software. Furthermore, we provide a repository on Github\footnote{The
  repository can be found on
  \url{https://github.com/mpbron/allib-chao-experiments}. The repository
  of \texttt{python-allib} can be found on
  \url{https://github.com/mpbron/allib}.} and ZENODO (see{[}3{]}) that
contains the scripts which the reader can use to reproduce our results.

\bookmarksetup{startatroot}

\section{Results}\label{results}

\subsection{Comparing Sampling
strategies}\label{comparing-sampling-strategies}

In Table~\ref{tbl-work-comparison}, we show the performance of each
sampling strategy. The AutoTAR method globally outperforms the others,
which is to be expected as this method does not perform any additional
work to enable the estimation of the current recall. This is also a
result reported in earlier work (e.g.,{[}31{]}). Overall, AutoTAR has a
mean WSS@95 of 74.3 \% vs.~ours (Ensemble) of 63 \% and 45.5 \% for
AUTOSTOP. Our method does not outperform AutoTAR in terms of WSS@95.
This result was to be expected for our Ensemble method, because our
method uses Random Sampling for selecting approximately 20 \% of the
instances. Note that the reported standard deviation in
Table~\ref{tbl-work-comparison} indicates that some datasets are more
difficult than others. For example, AutoTAR only achieves a WSS@95 of
10.3 \% on the Moran dataset. Considering only the WSS@95 and WSS@100,
our method stays closer to the performance of AutoTAR than AUTOSTOP.

\begin{table}

\caption{\label{tbl-work-comparison}Comparison of Work Savings between
Active Learning strategies when using a Perfect Stopping criterion that
directly stops after the recall target has been achieved.}

\centering{

\centering
\begin{tabular}{llccc}
\toprule
\multicolumn{2}{c}{ } & \multicolumn{3}{c}{\textbf{Sampling Strategy}} \\
\cmidrule(l{3pt}r{3pt}){3-5}
\textbf{Target} &  & \textbf{AUTOSTOP} & \textbf{AutoTAR} & \textbf{Ensemble}\\
\midrule
\cellcolor{gray!15}{\textbf{95 \%}} & \cellcolor{gray!15}{\textbf{Effort (\%)}} & \cellcolor{gray!15}{51 ± 12} & \cellcolor{gray!15}{22 ± 16} & \cellcolor{gray!15}{33 ± 19}\\
\textbf{} & \textbf{WSS (\%)} & 45 ± 13 & 74 ± 16 & 63 ± 19\\
\cellcolor{gray!15}{\textbf{}} & \cellcolor{gray!15}{\textbf{loss-er}} & \cellcolor{gray!15}{0.09 ± 0.07} & \cellcolor{gray!15}{0.02 ± 0.05} & \cellcolor{gray!15}{0.05 ± 0.07}\\
\textbf{100 \%} & \textbf{Effort (\%)} & 66 ± 18 & 42 ± 29 & 52 ± 26\\
\cellcolor{gray!15}{\textbf{}} & \cellcolor{gray!15}{\textbf{WSS (\%)}} & \cellcolor{gray!15}{34 ± 18} & \cellcolor{gray!15}{58 ± 29} & \cellcolor{gray!15}{48 ± 26}\\
\textbf{} & \textbf{loss-er} & 0.14 ± 0.10 & 0.07 ± 0.10 & 0.09 ± 0.10\\
\bottomrule
\end{tabular}

}

\end{table}%

\subsection{Recall and Estimator
curves}\label{recall-and-estimator-curves}

In our experiments, the stopping criteria are called every ten review
decisions. If the stopping criterion uses an estimator, it will also
record the current point estimate and confidence interval. In a
real-world application, the system can plot these estimates together
with the recall statistics of the process. This plot can be an
informative aid to users in deciding whether or not to stop the review
process.

\begin{figure}

\begin{minipage}{0.50\linewidth}

\centering{

\includegraphics{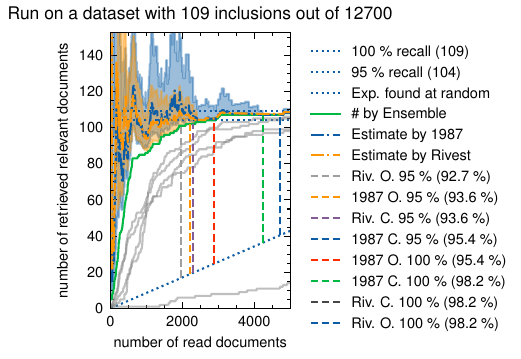}

}

\subcaption{\label{fig-recall-plot-1}Chao -- CLEF2017-CD011548 dataset
(ours)}

\end{minipage}%
\begin{minipage}{0.50\linewidth}

\centering{

\includegraphics{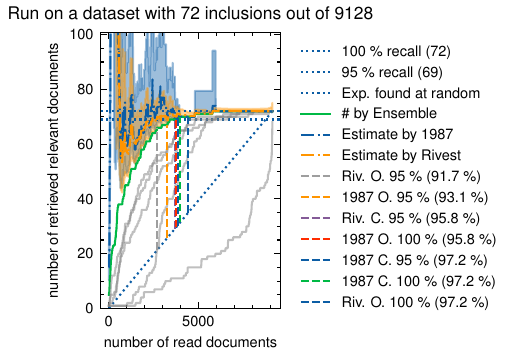}

}

\subcaption{\label{fig-recall-plot-2}Chao -- Van Dis dataset (ours)}

\end{minipage}%
\newline
\begin{minipage}{0.50\linewidth}

\centering{

\includegraphics{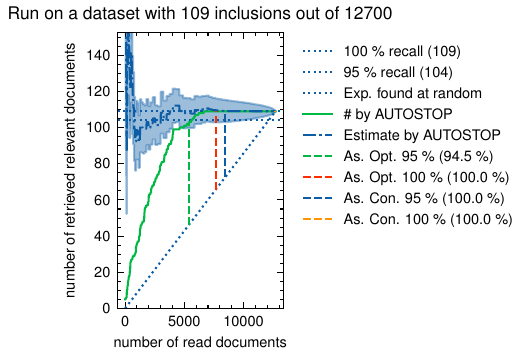}

}

\subcaption{\label{fig-recall-plot-3}AUTOSTOP -- CLEF2017-CD011548
dataset}

\end{minipage}%
\begin{minipage}{0.50\linewidth}

\centering{

\includegraphics{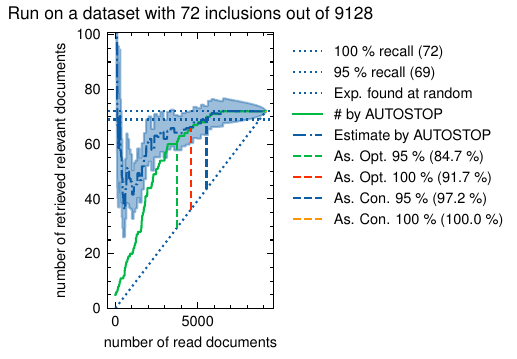}

}

\subcaption{\label{fig-recall-plot-4}AUTOSTOP -- Van Dis dataset}

\end{minipage}%
\newline
\begin{minipage}{0.50\linewidth}

\centering{

\includegraphics{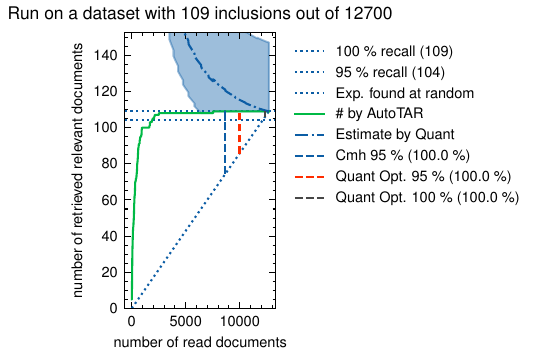}

}

\subcaption{\label{fig-recall-plot-5}Quant \& CMH-- CLEF2017-CD011548
dataset}

\end{minipage}%
\begin{minipage}{0.50\linewidth}

\centering{

\includegraphics{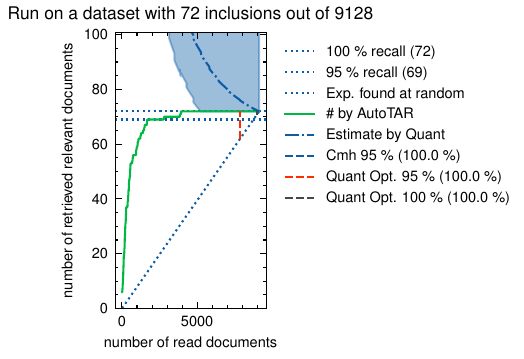}

}

\subcaption{\label{fig-recall-plot-6}Quant \& CMH -- Van Dis dataset}

\end{minipage}%

\caption{\label{fig-recall-plot}Recall curves for two datasets. The
dashed blue diagonal line shows how many documents would have been found
at random. The horizontal lines show the 95 and 100 \% recall targets.
The vertical dashed lines show when the stopping criteria have been
triggered. The ribbons around the estimates show their confidence
intervals.}

\end{figure}

In Figure~\ref{fig-recall-plot}, we show and compare the estimates and
stopping points of our method, AUTOSTOP, Quant, and CMH for runs of two
datasets. These methods allow the specification of recall targets,
however the CMH method does not provide an estimate on the current level
of recall or the number of relevant doucments. Note that for our method
(Chao), we show the results of the individual committee members from
within the ensemble. These members are displayed in light gray.

In Figure~\ref{fig-recall-plot-1}, our estimator fluctuates drastically
during the start of the process. When more documents are found by
multiple committee members, the estimates become more stable although
new documents keep being discovered. A similar behavior is visible for
AUTOSTOP in Figure~\ref{fig-recall-plot-3}. The Quant rule {[}41{]}
overestimates the number of relevant documents for an extended period.
Because of the large CI, Quant's Conservative recall criteria are never
triggered for the high recall targets. The \emph{Chao (Rivest) -
Conservative} 100 \% target is not triggered for our method on the Van
Dis dataset. AUTOSTOP's criterion is only triggered a few documents
before all documents are exhausted (holds for both datasets). Note that
for the runs displayed in Figure~\ref{fig-recall-plot}, the seed sets
for a dataset are kept the same among all methods, allowing a fair
comparison. For these runs, our methods need less reader effort than
AUTOSTOP. Furthermore, the AUTOSTOP requires less effort than the CMH
and Quant methods, eventhough AutoTAR is more efficient in retrieving
all relevant documents than AUTOSTOP.

\subsection{Criteria with recall
targets}\label{criteria-with-recall-targets}

In this section, we compare the criteria with recall targets over all
datasets. We will consider the result metrics for each time the stop
criteria were triggered. For each of the 94 datasets in our collection,
we have 30 runs with for each run a different seed sets. This results in
2820 runs per method. In Table~\ref{tbl-est-comparison-cons} and
Table~\ref{tbl-est-comparison-opt}, the results for all seven metrics
are given for all methods and, if applicable, for various recall
targets. We report the mean and standard deviation (if applicable) for
each score, enabling the reader to assess the dispersion of each score.
We discuss the results of the Conservative and Optimistic criteria
separately. Moreover, we report the results of the CMH method {[}8{]},
which does not provide estimates but allows the specification of a
recall target. We list the standoff version as in
Table~\ref{tbl-est-comparison-cmh}. We will discuss several recall
targets, but we will mainly focus on the high 95 \% and 100 \% recall
targets.

\subsubsection{Conservative methods}\label{conservative-methods}

\begin{table}

\caption{\label{tbl-est-comparison-cons}This table shows all metrics
recorded for each Conservative criterion. Note that for the Effort,
Recall, WSS, loss\(_\textrm{er}\) and Error, the mean and standard
deviation are reported.}

\centering{

\centering
\begin{tabular}{llcccc}
\toprule
\multicolumn{2}{c}{ } & \multicolumn{4}{c}{\textbf{Conservative Methods}} \\
\cmidrule(l{3pt}r{3pt}){3-6}
\textbf{Target} &  & \textbf{AUTOSTOP} & \textbf{Chao (1987)} & \textbf{Chao (Rivest)} & \textbf{Quant}\\
\midrule
\cellcolor{gray!15}{\textbf{70 \%}} & \cellcolor{gray!15}{\textbf{Effort (\%)}} & \cellcolor{gray!15}{28 ± 6} & \cellcolor{gray!15}{17 ± 13} & \cellcolor{gray!15}{17 ± 12} & \cellcolor{gray!15}{83 ± 23}\\
\textbf{} & \textbf{Recall (\%)} & 74 ± 9 & 74 ± 14 & 75 ± 14 & 100 ± 1\\
\cellcolor{gray!15}{\textbf{}} & \cellcolor{gray!15}{\textbf{WSS (\%)}} & \cellcolor{gray!15}{46 ± 12} & \cellcolor{gray!15}{57 ± 16} & \cellcolor{gray!15}{58 ± 16} & \cellcolor{gray!15}{16 ± 22}\\
\textbf{} & \textbf{loss-er} & 0.11 ± 0.06 & 0.10 ± 0.09 & 0.10 ± 0.08 & 0.33 ± 0.25\\
\cellcolor{gray!15}{\textbf{}} & \cellcolor{gray!15}{\textbf{Error (\%)}} & \cellcolor{gray!15}{11 ± 10} & \cellcolor{gray!15}{18 ± 12} & \cellcolor{gray!15}{17 ± 12} & \cellcolor{gray!15}{42 ± 2}\\
\textbf{} & \textbf{Target met (\%)} & 66 & 64 & 65 & 100\\
\cellcolor{gray!15}{\textbf{}} & \cellcolor{gray!15}{\textbf{Triggered (\%)}} & \cellcolor{gray!15}{100} & \cellcolor{gray!15}{100} & \cellcolor{gray!15}{100} & \cellcolor{gray!15}{44}\\
\textbf{80 \%} & \textbf{Effort (\%)} & 37 ± 6 & 22 ± 16 & 21 ± 14 & 93 ± 16\\
\cellcolor{gray!15}{\textbf{}} & \cellcolor{gray!15}{\textbf{Recall (\%)}} & \cellcolor{gray!15}{85 ± 8} & \cellcolor{gray!15}{83 ± 11} & \cellcolor{gray!15}{82 ± 11} & \cellcolor{gray!15}{100 ± 0}\\
\textbf{} & \textbf{WSS (\%)} & 47 ± 10 & 61 ± 16 & 61 ± 16 & 7 ± 15\\
\cellcolor{gray!15}{\textbf{}} & \cellcolor{gray!15}{\textbf{loss-er}} & \cellcolor{gray!15}{0.09 ± 0.05} & \cellcolor{gray!15}{0.07 ± 0.06} & \cellcolor{gray!15}{0.07 ± 0.06} & \cellcolor{gray!15}{0.36 ± 0.23}\\
\textbf{} & \textbf{Error (\%)} & 9 ± 7 & 12 ± 8 & 11 ± 8 & 25 ± 0\\
\cellcolor{gray!15}{\textbf{}} & \cellcolor{gray!15}{\textbf{Target met (\%)}} & \cellcolor{gray!15}{76} & \cellcolor{gray!15}{65} & \cellcolor{gray!15}{62} & \cellcolor{gray!15}{100}\\
\textbf{} & \textbf{Triggered (\%)} & 100 & 100 & 100 & 20\\
\cellcolor{gray!15}{\textbf{90 \%}} & \cellcolor{gray!15}{\textbf{Effort (\%)}} & \cellcolor{gray!15}{56 ± 11} & \cellcolor{gray!15}{31 ± 20} & \cellcolor{gray!15}{29 ± 18} & \cellcolor{gray!15}{99 ± 3}\\
\textbf{} & \textbf{Recall (\%)} & 94.8 ± 5.3 & 92.1 ± 6.7 & 90.7 ± 7.1 & 100.0 ± 0.0\\
\cellcolor{gray!15}{\textbf{}} & \cellcolor{gray!15}{\textbf{WSS (\%)}} & \cellcolor{gray!15}{39 ± 11} & \cellcolor{gray!15}{61 ± 19} & \cellcolor{gray!15}{62 ± 17} & \cellcolor{gray!15}{1 ± 3}\\
\textbf{} & \textbf{loss-er} & 0.15 ± 0.14 & 0.06 ± 0.08 & 0.06 ± 0.07 & 0.36 ± 0.22\\
\cellcolor{gray!15}{\textbf{}} & \cellcolor{gray!15}{\textbf{Error (\%)}} & \cellcolor{gray!15}{6.6 ± 4.4} & \cellcolor{gray!15}{6.4 ± 4.6} & \cellcolor{gray!15}{6.4 ± 4.8} & \cellcolor{gray!15}{11.1 ± 0.0}\\
\textbf{} & \textbf{Target met (\%)} & 86 & 68 & 59 & 100\\
\cellcolor{gray!15}{\textbf{}} & \cellcolor{gray!15}{\textbf{Triggered (\%)}} & \cellcolor{gray!15}{100} & \cellcolor{gray!15}{100} & \cellcolor{gray!15}{100} & \cellcolor{gray!15}{5.3}\\
\textbf{95 \%} & \textbf{Effort (\%)} & 74 ± 14 & 40 ± 23 & 41 ± 25 & 100 ± 0\\
\cellcolor{gray!15}{\textbf{}} & \cellcolor{gray!15}{\textbf{Recall (\%)}} & \cellcolor{gray!15}{98.2 ± 3.3} & \cellcolor{gray!15}{95.9 ± 5.0} & \cellcolor{gray!15}{95.3 ± 4.2} & \cellcolor{gray!15}{100.0 ± 0.0}\\
\textbf{} & \textbf{WSS (\%)} & 24 ± 12 & 56 ± 22 & 55 ± 24 & 0 ± 0\\
\cellcolor{gray!15}{\textbf{}} & \cellcolor{gray!15}{\textbf{loss-er}} & \cellcolor{gray!15}{0.25 ± 0.21} & \cellcolor{gray!15}{0.08 ± 0.10} & \cellcolor{gray!15}{0.11 ± 0.19} & \cellcolor{gray!15}{0.37 ± 0.22}\\
\textbf{} & \textbf{Error (\%)} & 4.20 ± 2.39 & 3.88 ± 3.63 & 3.62 ± 2.51 & 5.26 ± 0.00\\
\cellcolor{gray!15}{\textbf{}} & \cellcolor{gray!15}{\textbf{Target met (\%)}} & \cellcolor{gray!15}{89} & \cellcolor{gray!15}{70} & \cellcolor{gray!15}{59} & \cellcolor{gray!15}{100}\\
\textbf{} & \textbf{Triggered (\%)} & 100 & 100 & 94 & 0\\
\cellcolor{gray!15}{\textbf{100 \%}} & \cellcolor{gray!15}{\textbf{Effort (\%)}} & \cellcolor{gray!15}{100 ± 1} & \cellcolor{gray!15}{52 ± 27} & \cellcolor{gray!15}{84 ± 26} & \cellcolor{gray!15}{100 ± 0}\\
\textbf{} & \textbf{Recall (\%)} & 100.00 ± 0.07 & 98.02 ± 4.35 & 99.47 ± 1.06 & 100.00 ± 0.00\\
\cellcolor{gray!15}{\textbf{}} & \cellcolor{gray!15}{\textbf{WSS (\%)}} & \cellcolor{gray!15}{0 ± 0} & \cellcolor{gray!15}{46 ± 26} & \cellcolor{gray!15}{15 ± 25} & \cellcolor{gray!15}{0 ± 0}\\
\textbf{} & \textbf{loss-er} & 0.36 ± 0.22 & 0.10 ± 0.10 & 0.34 ± 0.25 & 0.37 ± 0.22\\
\cellcolor{gray!15}{\textbf{}} & \cellcolor{gray!15}{\textbf{Error (\%)}} & \cellcolor{gray!15}{0.00 ± 0.07} & \cellcolor{gray!15}{1.98 ± 4.35} & \cellcolor{gray!15}{0.53 ± 1.06} & \cellcolor{gray!15}{0.00 ± 0.00}\\
\textbf{} & \textbf{Target met (\%)} & 100 & 53 & 70 & 100\\
\cellcolor{gray!15}{\textbf{}} & \cellcolor{gray!15}{\textbf{Triggered (\%)}} & \cellcolor{gray!15}{56} & \cellcolor{gray!15}{99} & \cellcolor{gray!15}{33} & \cellcolor{gray!15}{0}\\
\bottomrule
\end{tabular}

}

\end{table}%

\begin{figure}

\begin{minipage}{0.43\linewidth}

\centering{

\includegraphics{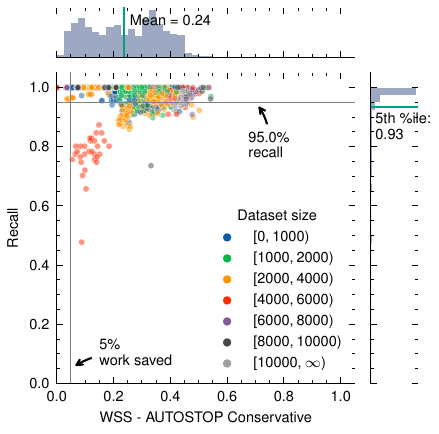}

}

\subcaption{\label{fig-cmhplots-cons-95-1}AUTOSTOP}

\end{minipage}%
\begin{minipage}{0.14\linewidth}
~\end{minipage}%
\begin{minipage}{0.43\linewidth}

\centering{

\includegraphics{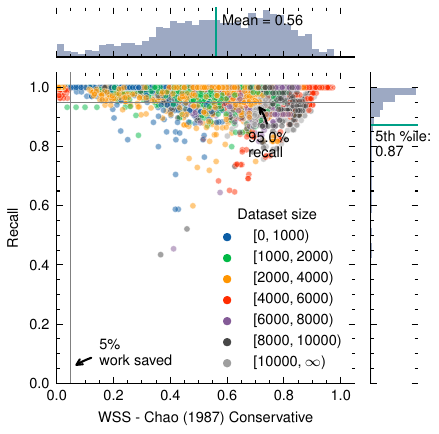}

}

\subcaption{\label{fig-cmhplots-cons-95-2}Chao (1987)}

\end{minipage}%
\newline
\begin{minipage}{0.43\linewidth}

\centering{

\includegraphics{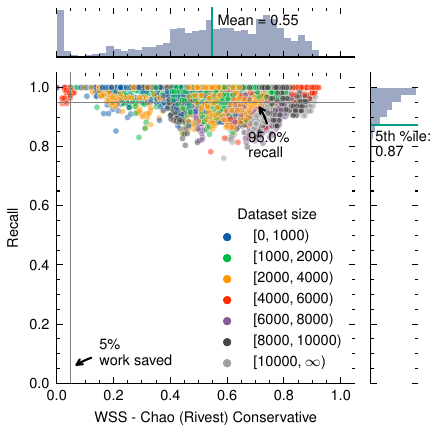}

}

\subcaption{\label{fig-cmhplots-cons-95-3}Chao (Rivest)}

\end{minipage}%
\begin{minipage}{0.14\linewidth}
~\end{minipage}%
\begin{minipage}{0.43\linewidth}

\centering{

\includegraphics{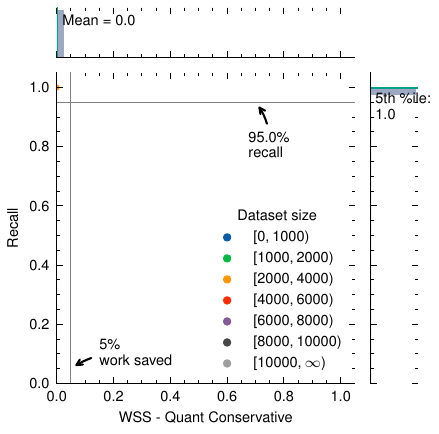}

}

\subcaption{\label{fig-cmhplots-cons-95-4}Quant}

\end{minipage}%

\caption{\label{fig-cmhplots-cons-95}These figures display the results
of all runs on all datasets in terms of Work Saved over Sampling and
Recall for the Conservative Stopping criteria with a 95 \% recall
target. The colors are indicative of the dataset size. The outer graphs
show the overall distribution of WSS and Recall. For each method, the
mean WSS and the 5th percentile of the recall are shown.}

\end{figure}%

\begin{figure}

\begin{minipage}{0.43\linewidth}

\centering{

\includegraphics{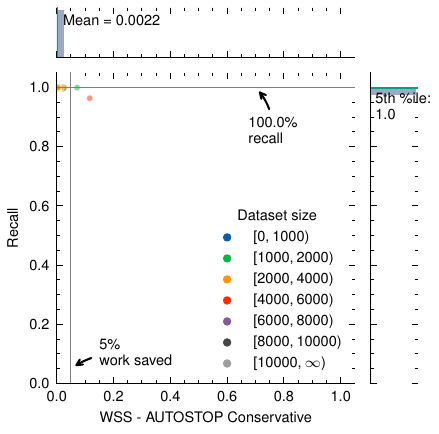}

}

\subcaption{\label{fig-cmhplots-cons-100-1}AUTOSTOP}

\end{minipage}%
\begin{minipage}{0.14\linewidth}
~\end{minipage}%
\begin{minipage}{0.43\linewidth}

\centering{

\includegraphics{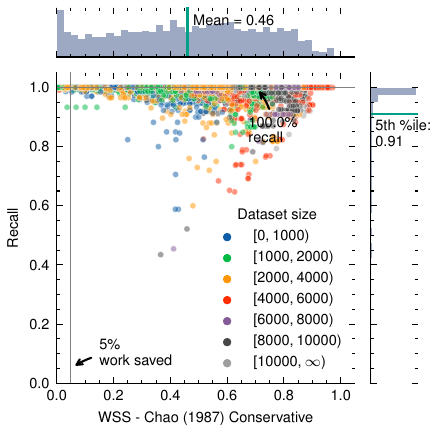}

}

\subcaption{\label{fig-cmhplots-cons-100-2}Chao (1987)}

\end{minipage}%
\newline
\begin{minipage}{0.43\linewidth}

\centering{

\includegraphics{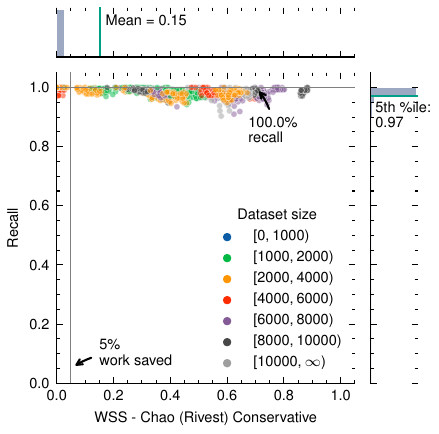}

}

\subcaption{\label{fig-cmhplots-cons-100-3}Chao (Rivest)}

\end{minipage}%
\begin{minipage}{0.14\linewidth}
~\end{minipage}%
\begin{minipage}{0.43\linewidth}

\centering{

\includegraphics{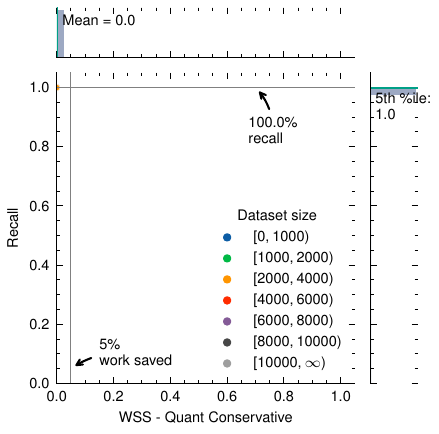}

}

\subcaption{\label{fig-cmhplots-cons-100-4}Quant}

\end{minipage}%

\caption{\label{fig-cmhplots-cons-100}These figures display the results
of all runs on all datasets in terms of Work Saved over Sampling and
Recall for the Conservative Stopping criteria with a 100 \% recall
target. The colors are indicative of the dataset size. The outer graphs
show the overall distribution of WSS and Recall. For each method, the
mean WSS and the 5th percentile of the recall are shown.}

\end{figure}

In Figure~\ref{fig-cmhplots-cons-95} and
Figure~\ref{fig-cmhplots-cons-100}, the results of all the runs with a
95 \% and 100 \% recall target are displayed. The analysis and figures
are based on work performed in {[}8{]}, in which the authors performed
and presented a similar analysis.

When comparing the Conservative methods for 95 \%, we can see that while
the tail of the distribution of the \emph{Chao (1987)} method of the
recall is 5.91 percent points lower than \emph{AUTOSTOP'}s recall, our
WSS is 32.46 points higher. Our method's recall vs.~work savings
trade-off is slightly more leaned towards the latter. The results of the
two Chao versions are similar but slightly in favor of Chao's original
estimator. Note that for \emph{Chao (Rivest)}, there are many runs for
which the work savings are below 5 \%. This is still the case for
\emph{Chao (1987)}; however, it is less pronounced.

For the 100 \% target recall, the differences between the criteria are
more pronounced. For instance, the AUTOSTOP criterion is not triggered
in many runs, and the work savings are minimal for the few runs it is
triggered. This is even more the case for the Quant Rule, as it is never
triggered. There is a large difference between the Work Savings between
\emph{Chao (Rivest)} and \emph{Chao (1987)} for the 100 \% criterion.
This is because the \emph{Chao (Rivest)} can also provide a CI for when
\(f_1\) and \(f_2\) both become zero, whereas \emph{Chao (1987)} cannot.
Also, when \(\hat{N} = n\), the upper bound of the CI is by definition
\(n\) (see Equation~\ref{eq-chao-ci1}).

For all conservative criteria, it holds that when the upper bound of the
CI \(\hat{N}_{\textrm{sup}} > n\), the stopping criteria for 100 \%
recall cannot be triggered (unless \(\hat{R}\) is rounded up to 100 \%).
For several datasets, the \emph{Chao (1987) - Conservative} estimates
are not triggered during the run due to the fact that the CI is still
not small enough (e.g., visible in Figure~\ref{fig-recall-plot-2}).
However, for the runs that it is triggered, this still results in a mean
recall of 46.05 \%.

Considering the lower recall targets (e.g., 70 \% and 80\%), our
method's mean recall (as presented in
Table~\ref{tbl-est-comparison-cons}) is slightly higher than the target
recall. However, the standard deviation and error rates of all estimator
methods decrease as the recall target increases. This result was also
reported in {[}31{]}. The Quant Conservative method is not often
triggered; this results in a very high recall, even for the lower recall
targets. The percentage of times this criterion is triggered tends to
zero as the recall target increases.

Regarding reliability for the high recall targets, our method does not
achieve its recall target as often as AUTOSTOP. However, our method is
close, as reported by the low error rate. Given the higher work savings
and the mean recall compared to the other estimator methods, especially
the \emph{Chao (1987)} method provides a good alternative to the
AUTOSTOP Criterion. This is especially true for the 100 \% criteria. As
the \emph{Chao (1987)} criterion is triggered 99.37 \% of the time
vs.~55.93 \% for AUTOSTOP.

\subsubsection{Optimistic methods}\label{optimistic-methods}

\begin{figure}

\begin{minipage}{0.43\linewidth}

\centering{

\includegraphics{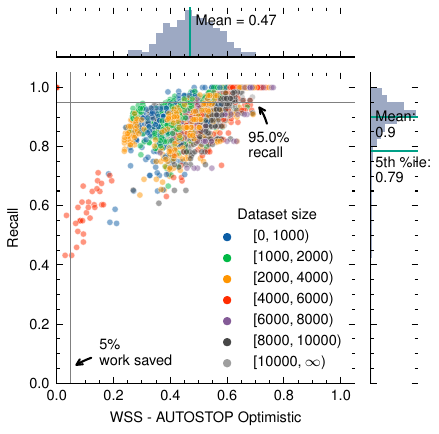}

}

\subcaption{\label{fig-cmhplots-opt-95-1}AUTOSTOP}

\end{minipage}%
\begin{minipage}{0.14\linewidth}
~\end{minipage}%
\begin{minipage}{0.43\linewidth}

\centering{

\includegraphics{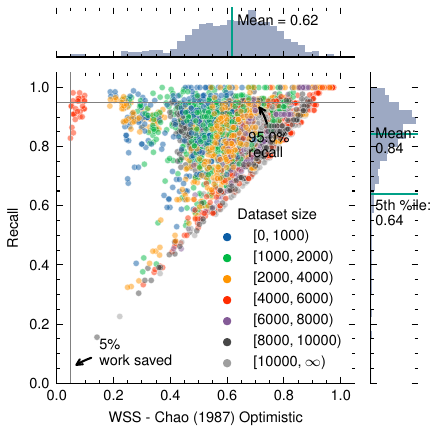}

}

\subcaption{\label{fig-cmhplots-opt-95-2}Chao (1987)}

\end{minipage}%
\newline
\begin{minipage}{0.43\linewidth}

\centering{

\includegraphics{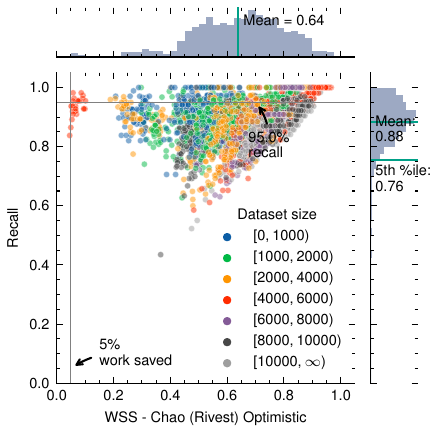}

}

\subcaption{\label{fig-cmhplots-opt-95-3}Chao (Rivest)}

\end{minipage}%
\begin{minipage}{0.14\linewidth}
~\end{minipage}%
\begin{minipage}{0.43\linewidth}

\centering{

\includegraphics{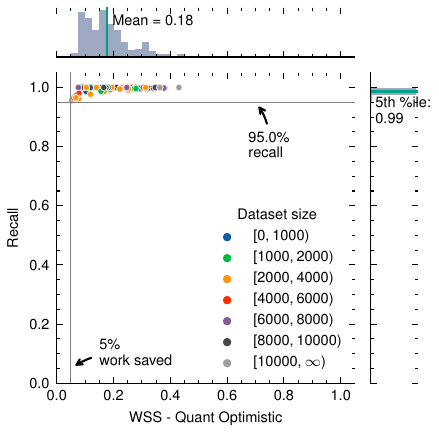}

}

\subcaption{\label{fig-cmhplots-opt-95-4}Quant}

\end{minipage}%

\caption{\label{fig-cmhplots-opt-95}These figures display the results of
all runs on all datasets in terms of Work Saved over Sampling and Recall
for the Optimistic Stopping criteria with a 95 \% recall target. The
colors are indicative of the dataset size. The outer graphs show the
overall distribution of WSS and Recall. For each method, the mean WSS
and the 5th percentile of the recall are shown.}

\end{figure}

For the optimistic methods, the results are presented in
Table~\ref{tbl-est-comparison-opt}. In Figure~\ref{fig-cmhplots-opt-95},
the results of all the estimator methods are displayed. The achieved
recall by our stopping criteria is lower than AUTOSTOP reaches; this is
especially visible through the 5th percentile of the recall scores,
which is 3.03 percent points lower for \emph{Chao (Rivest)}. There is
also a significant difference between our criteria, as the 5th
percentile is 11.59 points higher for Rivest's Poisson Regression
version. Our results show that \emph{Chao (Rivest)} outperforms
\emph{Chao (1987)} for all recall targets in terms of recall. For the
100 \% recall target, both our estimators improve over AUTOSTOP. For
\emph{Chao (1987),} the recall distributions are similar, while the work
savings are improved. For \emph{Chao (Rivest),} the 5th percentile of
the recall scores lies at 94.5 \%, which is a large improvement over
AUTOSTOP (86.21 \%). However, AUTOSTOP provides a slightly higher recall
than \emph{Chao (Rivest)} for the lower recall targets. A trait that is
visible for all Optimistic criteria is that all methods tend to
underestimate the number of relevant documents. However, the Recall
Error decreases as the recall target becomes higher.

The Quant method overestimates the number of relevant documents for all
recall targets. While the Quant method meets the recall target in nearly
all cases, the error in recall prediction is very high. This is visible
in the results for the 70 \% recall target: for nearly all runs, the
method stops at a point where (almost) all relevant documents from the
dataset are retrieved.

\begin{table}

\caption{\label{tbl-est-comparison-opt}This table shows all metrics
recorded for each Optimistic criterion, as well as the CMH method. Note
that for the Effort, Recall, WSS, loss\(_\textrm{er}\) and Error, the
mean and standard deviation are reported.}

\centering{

\centering
\begin{tabular}{llcccc}
\toprule
\multicolumn{2}{c}{ } & \multicolumn{4}{c}{\textbf{Optimistic Methods}} \\
\cmidrule(l{3pt}r{3pt}){3-6}
\textbf{Target} &  & \textbf{AUTOSTOP} & \textbf{Chao (1987)} & \textbf{Chao (Rivest)} & \textbf{Quant}\\
\midrule
\cellcolor{gray!15}{\textbf{70 \%}} & \cellcolor{gray!15}{\textbf{Effort (\%)}} & \cellcolor{gray!15}{23 ± 7} & \cellcolor{gray!15}{11 ± 9} & \cellcolor{gray!15}{12 ± 10} & \cellcolor{gray!15}{45 ± 10}\\
\textbf{} & \textbf{Recall (\%)} & 66 ± 9 & 57 ± 17 & 63 ± 16 & 98 ± 3\\
\cellcolor{gray!15}{\textbf{}} & \cellcolor{gray!15}{\textbf{WSS (\%)}} & \cellcolor{gray!15}{42 ± 11} & \cellcolor{gray!15}{47 ± 16} & \cellcolor{gray!15}{51 ± 16} & \cellcolor{gray!15}{53 ± 10}\\
\textbf{} & \textbf{loss-er} & 0.15 ± 0.07 & 0.22 ± 0.14 & 0.17 ± 0.12 & 0.10 ± 0.08\\
\cellcolor{gray!15}{\textbf{}} & \cellcolor{gray!15}{\textbf{Error (\%)}} & \cellcolor{gray!15}{11 ± 10} & \cellcolor{gray!15}{24 ± 17} & \cellcolor{gray!15}{20 ± 15} & \cellcolor{gray!15}{40 ± 5}\\
\textbf{} & \textbf{Target met (\%)} & 27 & 24 & 33 & 100\\
\cellcolor{gray!15}{\textbf{}} & \cellcolor{gray!15}{\textbf{Triggered (\%)}} & \cellcolor{gray!15}{100} & \cellcolor{gray!15}{100} & \cellcolor{gray!15}{100} & \cellcolor{gray!15}{\vphantom{3} 100}\\
\textbf{80 \%} & \textbf{Effort (\%)} & 30 ± 8 & 13 ± 11 & 15 ± 11 & 57 ± 10\\
\cellcolor{gray!15}{\textbf{}} & \cellcolor{gray!15}{\textbf{Recall (\%)}} & \cellcolor{gray!15}{76 ± 9} & \cellcolor{gray!15}{65 ± 16} & \cellcolor{gray!15}{71 ± 14} & \cellcolor{gray!15}{99 ± 2}\\
\textbf{} & \textbf{WSS (\%)} & 46 ± 11 & 52 ± 16 & 56 ± 15 & 43 ± 11\\
\cellcolor{gray!15}{\textbf{}} & \cellcolor{gray!15}{\textbf{loss-er}} & \cellcolor{gray!15}{0.10 ± 0.06} & \cellcolor{gray!15}{0.16 ± 0.12} & \cellcolor{gray!15}{0.12 ± 0.08} & \cellcolor{gray!15}{0.14 ± 0.12}\\
\textbf{} & \textbf{Error (\%)} & 9 ± 8 & 22 ± 16 & 16 ± 12 & 24 ± 2\\
\cellcolor{gray!15}{\textbf{}} & \cellcolor{gray!15}{\textbf{Target met (\%)}} & \cellcolor{gray!15}{28} & \cellcolor{gray!15}{17} & \cellcolor{gray!15}{27} & \cellcolor{gray!15}{100}\\
\textbf{} & \textbf{Triggered (\%)} & 100 & 100 & 100 & \vphantom{2} 100\\
\cellcolor{gray!15}{\textbf{90 \%}} & \cellcolor{gray!15}{\textbf{Effort (\%)}} & \cellcolor{gray!15}{38 ± 8} & \cellcolor{gray!15}{18 ± 14} & \cellcolor{gray!15}{20 ± 14} & \cellcolor{gray!15}{72 ± 10}\\
\textbf{} & \textbf{Recall (\%)} & 86 ± 7 & 77 ± 13 & 82 ± 10 & 100 ± 1\\
\cellcolor{gray!15}{\textbf{}} & \cellcolor{gray!15}{\textbf{WSS (\%)}} & \cellcolor{gray!15}{48 ± 11} & \cellcolor{gray!15}{59 ± 15} & \cellcolor{gray!15}{62 ± 15} & \cellcolor{gray!15}{28 ± 10}\\
\textbf{} & \textbf{loss-er} & 0.08 ± 0.05 & 0.09 ± 0.08 & 0.06 ± 0.05 & 0.22 ± 0.16\\
\cellcolor{gray!15}{\textbf{}} & \cellcolor{gray!15}{\textbf{Error (\%)}} & \cellcolor{gray!15}{7 ± 7} & \cellcolor{gray!15}{16 ± 13} & \cellcolor{gray!15}{11 ± 9} & \cellcolor{gray!15}{11 ± 1}\\
\textbf{} & \textbf{Target met (\%)} & 24 & 12 & 21 & 100\\
\cellcolor{gray!15}{\textbf{}} & \cellcolor{gray!15}{\textbf{Triggered (\%)}} & \cellcolor{gray!15}{100} & \cellcolor{gray!15}{100} & \cellcolor{gray!15}{100} & \cellcolor{gray!15}{\vphantom{1} 100}\\
\textbf{95 \%} & \textbf{Effort (\%)} & 43 ± 9 & 23 ± 17 & 25 ± 16 & 82 ± 8\\
\cellcolor{gray!15}{\textbf{}} & \cellcolor{gray!15}{\textbf{Recall (\%)}} & \cellcolor{gray!15}{90 ± 7} & \cellcolor{gray!15}{84 ± 11} & \cellcolor{gray!15}{88 ± 7} & \cellcolor{gray!15}{100 ± 1}\\
\textbf{} & \textbf{WSS (\%)} & 47 ± 10 & 62 ± 16 & 64 ± 16 & 18 ± 8\\
\cellcolor{gray!15}{\textbf{}} & \cellcolor{gray!15}{\textbf{loss-er}} & \cellcolor{gray!15}{0.08 ± 0.05} & \cellcolor{gray!15}{0.06 ± 0.06} & \cellcolor{gray!15}{0.05 ± 0.04} & \cellcolor{gray!15}{0.27 ± 0.19}\\
\textbf{} & \textbf{Error (\%)} & 6 ± 6 & 12 ± 11 & 8 ± 6 & 5 ± 1\\
\cellcolor{gray!15}{\textbf{}} & \cellcolor{gray!15}{\textbf{Target met (\%)}} & \cellcolor{gray!15}{16} & \cellcolor{gray!15}{10} & \cellcolor{gray!15}{19} & \cellcolor{gray!15}{100}\\
\textbf{} & \textbf{Triggered (\%)} & 100 & 100 & 100 & 100\\
\cellcolor{gray!15}{\textbf{100 \%}} & \cellcolor{gray!15}{\textbf{Effort (\%)}} & \cellcolor{gray!15}{55 ± 11} & \cellcolor{gray!15}{39 ± 24} & \cellcolor{gray!15}{46 ± 21} & \cellcolor{gray!15}{98 ± 2}\\
\textbf{} & \textbf{Recall (\%)} & 95.9 ± 5.4 & 95.9 ± 5.5 & 98.3 ± 2.1 & 100.0 ± 0.1\\
\cellcolor{gray!15}{\textbf{}} & \cellcolor{gray!15}{\textbf{WSS (\%)}} & \cellcolor{gray!15}{41 ± 10} & \cellcolor{gray!15}{56 ± 22} & \cellcolor{gray!15}{52 ± 21} & \cellcolor{gray!15}{2 ± 2}\\
\textbf{} & \textbf{loss-er} & 0.10 ± 0.06 & 0.06 ± 0.07 & 0.09 ± 0.10 & 0.35 ± 0.22\\
\cellcolor{gray!15}{\textbf{}} & \cellcolor{gray!15}{\textbf{Error (\%)}} & \cellcolor{gray!15}{4.1 ± 5.4} & \cellcolor{gray!15}{4.1 ± 5.5} & \cellcolor{gray!15}{1.7 ± 2.1} & \cellcolor{gray!15}{0.0 ± 0.1}\\
\textbf{} & \textbf{Target met (\%)} & 20 & 23 & 39 & 98\\
\cellcolor{gray!15}{\textbf{}} & \cellcolor{gray!15}{\textbf{Triggered (\%)}} & \cellcolor{gray!15}{100} & \cellcolor{gray!15}{100} & \cellcolor{gray!15}{100} & \cellcolor{gray!15}{99}\\
\bottomrule
\end{tabular}

}

\end{table}%

\begin{figure}

\begin{minipage}{0.43\linewidth}

\centering{

\includegraphics{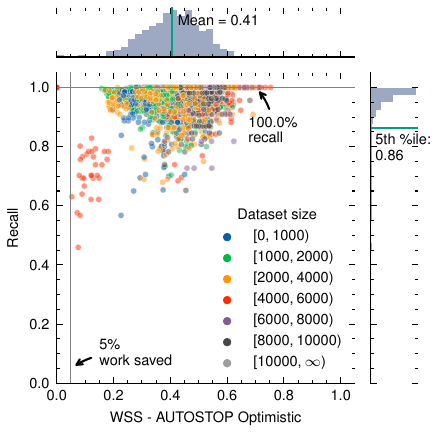}

}

\subcaption{\label{fig-cmhplots-opt-100-1}AUTOSTOP}

\end{minipage}%
\begin{minipage}{0.14\linewidth}
~\end{minipage}%
\begin{minipage}{0.43\linewidth}

\centering{

\includegraphics{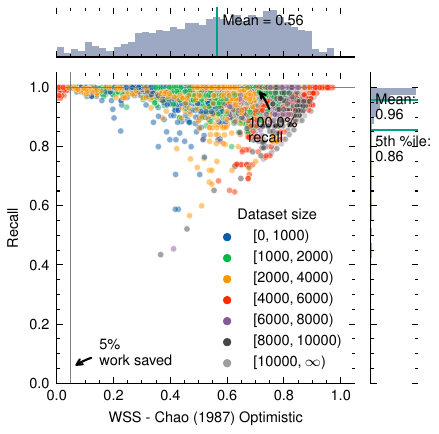}

}

\subcaption{\label{fig-cmhplots-opt-100-2}Chao (1987)}

\end{minipage}%
\newline
\begin{minipage}{0.43\linewidth}

\centering{

\includegraphics{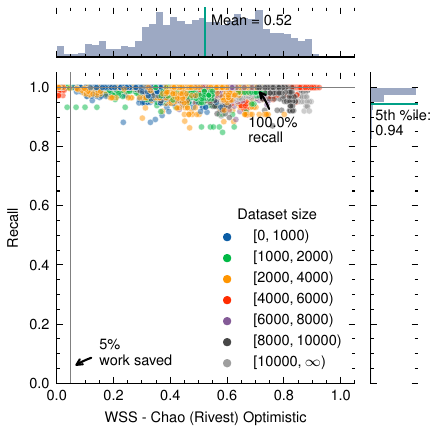}

}

\subcaption{\label{fig-cmhplots-opt-100-3}Chao (Rivest)}

\end{minipage}%
\begin{minipage}{0.14\linewidth}
~\end{minipage}%
\begin{minipage}{0.43\linewidth}

\centering{

\includegraphics{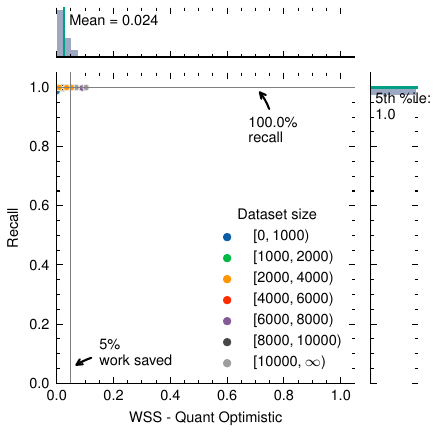}

}

\subcaption{\label{fig-cmhplots-opt-100-4}Quant}

\end{minipage}%

\caption{\label{fig-cmhplots-opt-100}These figures display the results
of all runs on all datasets in terms of Work Saved over Sampling and
Recall for the Optimistic Stopping criteria with a 100 \% recall target.
The colors are indicative of the dataset size. The outer graphs show the
overall distribution of WSS and Recall. For each method, the mean WSS
and the 5th percentile of the recall are shown.}

\end{figure}%

\subsubsection{CMH method (Standoff
version)}\label{cmh-method-standoff-version}

\begin{figure}

\begin{minipage}{0.43\linewidth}

\centering{

\includegraphics{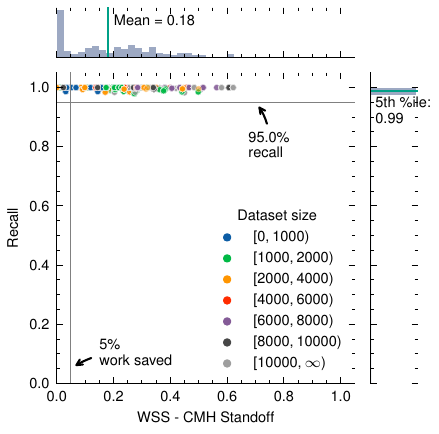}

}

\subcaption{\label{fig-cmhplots-cmh-1}CMH - 95 \%}

\end{minipage}%
\begin{minipage}{0.14\linewidth}
~\end{minipage}%
\begin{minipage}{0.43\linewidth}

\centering{

\includegraphics{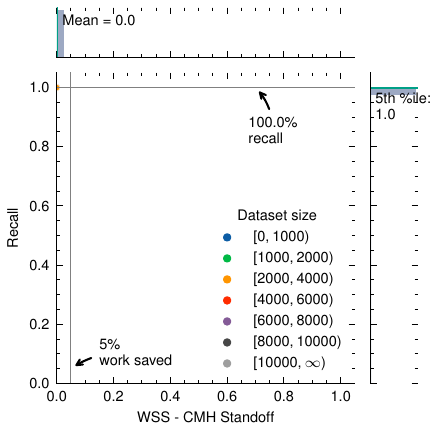}

}

\subcaption{\label{fig-cmhplots-cmh-2}CMH - 100 \%}

\end{minipage}%

\caption{\label{fig-cmhplots-cmh}These figures display the results of
all runs on all datasets in terms of Work Saved over Sampling and Recall
for the CMH criteria.}

\end{figure}

The results for the standoff version of the CMH method are displayed in
Table~\ref{tbl-est-comparison-cmh}. In Figure~\ref{fig-cmhplots-cmh},
the WSS and Recall scores for the CMH methods are displayed. In
Figure~\ref{fig-recall-plot-5} and Figure~\ref{fig-recall-plot-6}, the
results of the 95 \% target are shown for that particular run. The
results are similar to the results of the Quant Optimistic method,
although the percentage of runs that this criterion is triggered is
lower. This holds especially for the 100 \% and 95 \% targets: the 100
\% criterion is never triggered, and the 95 \% recall target is only
triggered 77.02 \% of the time. As the average recall for the 70 \%
recall target is already 60.1 \%, it is evident that this criterion
underestimates the recall for a long time.

\begin{table}

\caption{\label{tbl-est-comparison-cmh}This table shows all metrics
recorded for the CMH standoff method, for each of the studied recall
targets. Note that for the Effort, Recall, WSS, loss\(_\textrm{er}\) and
Error, the mean and standard deviation are reported.}

\centering{

\centering
\begin{tabular}{lccccc}
\toprule
\multicolumn{1}{c}{ } & \multicolumn{5}{c}{\textbf{CMH Standoff Method}} \\
\cmidrule(l{3pt}r{3pt}){2-6}
 & \textbf{70 \%} & \textbf{80 \%} & \textbf{90 \%} & \textbf{95 \%} & \textbf{100 \%}\\
\midrule
\cellcolor{gray!15}{\textbf{Effort (\%)}} & \cellcolor{gray!15}{38 ± 13} & \cellcolor{gray!15}{48 ± 15} & \cellcolor{gray!15}{65 ± 18} & \cellcolor{gray!15}{82 ± 16} & \cellcolor{gray!15}{100 ± 0}\\
\textbf{Recall (\%)} & 98.54 ± 2.23 & 99.18 ± 1.43 & 99.64 ± 0.74 & 99.87 ± 0.37 & 100.00 ± 0.00\\
\cellcolor{gray!15}{\textbf{WSS (\%)}} & \cellcolor{gray!15}{60 ± 14} & \cellcolor{gray!15}{51 ± 15} & \cellcolor{gray!15}{35 ± 18} & \cellcolor{gray!15}{18 ± 16} & \cellcolor{gray!15}{0 ± 0}\\
\textbf{loss-er} & 0.07 ± 0.07 & 0.11 ± 0.12 & 0.20 ± 0.21 & 0.30 ± 0.24 & 0.37 ± 0.22\\
\cellcolor{gray!15}{\textbf{Error (\%)}} & \cellcolor{gray!15}{41 ± 3} & \cellcolor{gray!15}{24 ± 2} & \cellcolor{gray!15}{11 ± 1} & \cellcolor{gray!15}{5 ± 0} & \cellcolor{gray!15}{0 ± 0}\\
\textbf{Target met (\%)} & 100 & 100 & 100 & 100 & 100\\
\cellcolor{gray!15}{\textbf{Triggered (\%)}} & \cellcolor{gray!15}{100} & \cellcolor{gray!15}{100} & \cellcolor{gray!15}{97} & \cellcolor{gray!15}{77} & \cellcolor{gray!15}{0}\\
\bottomrule
\end{tabular}

}

\end{table}%

\subsection{Hybrid methods}\label{hybrid-methods}

\begin{table}

\caption{\label{tbl-est-comparison-hybrid}This table shows all metrics
recorded for all hybrid methods. The target recall of the CMH method is
95 \%, the Target method does not allow the specification of a recall
target. Note that for the Effort, Recall, WSS, and loss\(_\textrm{er}\),
the mean and standard deviation are reported. Furthermore, we report the
rate various recall targets are achieved.}

\centering{

\centering
\begin{tabular}{lcc}
\toprule
\multicolumn{1}{c}{ } & \multicolumn{2}{c}{\textbf{Hybrid Methods}} \\
\cmidrule(l{3pt}r{3pt}){2-3}
 & \textbf{CMH} & \textbf{TARGET}\\
\midrule
\cellcolor{gray!15}{\textbf{Effort (\%)}} & \cellcolor{gray!15}{82 ± 16} & \cellcolor{gray!15}{38 ± 20}\\
\textbf{Recall (\%)} & 100 ± 0 & 92 ± 9\\
\cellcolor{gray!15}{\textbf{WSS (\%)}} & \cellcolor{gray!15}{18 ± 15} & \cellcolor{gray!15}{54 ± 18}\\
\textbf{loss-er} & 0.30 ± 0.24 & 0.10 ± 0.13\\
\cellcolor{gray!15}{\textbf{Target@70 (\%)}} & \cellcolor{gray!15}{100} & \cellcolor{gray!15}{97}\\
\textbf{Target@80 (\%)} & 100 & 90\\
\cellcolor{gray!15}{\textbf{Target@90 (\%)}} & \cellcolor{gray!15}{100} & \cellcolor{gray!15}{69}\\
\textbf{Target@95 (\%)} & 100 & 48\\
\cellcolor{gray!15}{\textbf{Target@100 (\%)}} & \cellcolor{gray!15}{80} & \cellcolor{gray!15}{19}\\
\textbf{Triggered (\%)} & 92 & 100\\
\bottomrule
\end{tabular}

}

\end{table}%

In this section, we study the results of two hybrid methods: the
CMH-Hybrid method {[}8{]} and the Target method {[}15{]}. While the CMH
method discussed in the previous section is a standoff method; this CMH
method is not a standoff method, as it consists of two phases: the first
phase consists of sampling using AutoTAR until the null hypothesis of
the hypergeometric test, as described in Section~\ref{sec-rl-cmh} is
rejected with \(\alpha = 0.5\) with a target recall of 95 \%. Then, the
method proceeds with screening through random sampling. The procedure is
stopped until the null hypothesis of the hypergeometric test is
rejected, but now with an \(\alpha = 0.05\). We studied the scenario
where the target recall is 95 \% for both phases (given the fact that
the 100 \% method is never triggered). The results of this test are
reported in Table~\ref{tbl-est-comparison-hybrid}. The results are
similar to the results reported in {[}8{]}, which indicated a recall
above 95 \% accompanied by a WSS of 17 \% (in our experiments 18.2\%),
on a different, but partially overlapping collection of test datasets.
The Target method reports a mean recall of 91.85 \%, with a WSS of 54.17
\%, which just differs 2.03 percent points from ours (our recall is
95.92 \% for \emph{Chao (1987) - Conservative}).

For the hybrid version of the CMH method, we can make the same remarks
as for the standoff version. The burden of rejecting the null hypothesis
of the Hypergeometric test is high, especially when \(\alpha = 0.05\).
This results in low work savings, although the achieved recall levels is
high.

\begin{figure}

\begin{minipage}{0.43\linewidth}

\centering{

\includegraphics{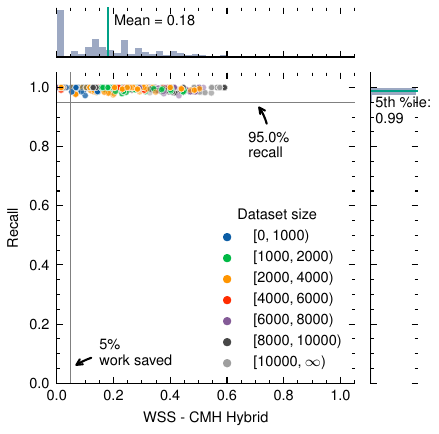}

}

\subcaption{\label{fig-cmhplots-hybrid-1}CMH}

\end{minipage}%
\begin{minipage}{0.14\linewidth}
~\end{minipage}%
\begin{minipage}{0.43\linewidth}

\centering{

\includegraphics{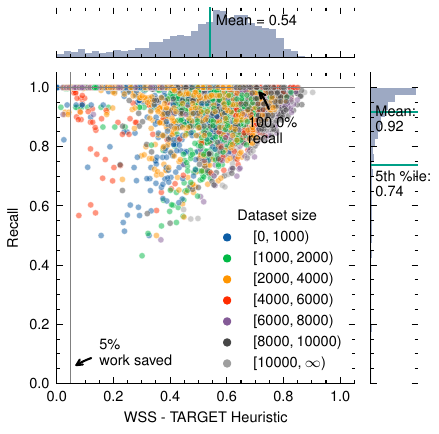}

}

\subcaption{\label{fig-cmhplots-hybrid-2}Target}

\end{minipage}%

\caption{\label{fig-cmhplots-hybrid}These figures display the results of
all runs on all datasets in terms of Work Saved over Sampling and Recall
for the hybrid stopping criteria.}

\end{figure}%

\subsection{Standoff Heuristics}\label{standoff-heuristics}

\begin{table}

\caption{\label{tbl-est-comparison-standoff}This table shows all metrics
recorded for all heuristic standoff methods. Note that for the Effort,
Recall, WSS, loss\(_\textrm{er}\), the mean and standard deviation are
reported. Furthermore, we report the rate at which various recall
targets are achieved, even though these are not specified by the
heuristics.}

\centering{

\centering
\begin{tabular}{lcccccc}
\toprule
\multicolumn{1}{c}{ } & \multicolumn{6}{c}{\textbf{Standoff Methods}} \\
\cmidrule(l{3pt}r{3pt}){2-7}
 & \textbf{Budget} & \textbf{Half} & \textbf{Knee} & \textbf{Rule2399} & \textbf{Stop200} & \textbf{Stop400}\\
\midrule
\cellcolor{gray!15}{\textbf{Effort (\%)}} & \cellcolor{gray!15}{35 ± 20} & \cellcolor{gray!15}{53 ± 2} & \cellcolor{gray!15}{82 ± 29} & \cellcolor{gray!15}{68 ± 32} & \cellcolor{gray!15}{41 ± 29} & \cellcolor{gray!15}{52 ± 31}\\
\textbf{Recall (\%)} & 97.04 ± 3.77 & 98.96 ± 3.15 & 99.25 ± 1.96 & 98.78 ± 4.74 & 98.13 ± 4.54 & 99.22 ± 2.23\\
\cellcolor{gray!15}{\textbf{WSS (\%)}} & \cellcolor{gray!15}{62 ± 19} & \cellcolor{gray!15}{46 ± 4} & \cellcolor{gray!15}{18 ± 28} & \cellcolor{gray!15}{31 ± 31} & \cellcolor{gray!15}{57 ± 28} & \cellcolor{gray!15}{47 ± 31}\\
\textbf{loss-er} & 0.08 ± 0.12 & 0.10 ± 0.06 & 0.34 ± 0.25 & 0.22 ± 0.22 & 0.07 ± 0.10 & 0.11 ± 0.14\\
\cellcolor{gray!15}{\textbf{Target@70 (\%)}} & \cellcolor{gray!15}{100} & \cellcolor{gray!15}{100} & \cellcolor{gray!15}{100} & \cellcolor{gray!15}{98} & \cellcolor{gray!15}{100} & \cellcolor{gray!15}{100}\\
\textbf{Target@80 (\%)} & 99 & 99 & 100 & 98 & 99 & 100\\
\cellcolor{gray!15}{\textbf{Target@90 (\%)}} & \cellcolor{gray!15}{95} & \cellcolor{gray!15}{98} & \cellcolor{gray!15}{100} & \cellcolor{gray!15}{98} & \cellcolor{gray!15}{97} & \cellcolor{gray!15}{99}\\
\textbf{Target@95 (\%)} & 76 & 95 & 93 & 95 & 89 & 96\\
\cellcolor{gray!15}{\textbf{Target@100 (\%)}} & \cellcolor{gray!15}{41} & \cellcolor{gray!15}{67} & \cellcolor{gray!15}{77} & \cellcolor{gray!15}{80} & \cellcolor{gray!15}{51} & \cellcolor{gray!15}{72}\\
\textbf{Triggered (\%)} & 100 & 100 & 34 & 60 & 94 & 85\\
\bottomrule
\end{tabular}

}

\end{table}%

\begin{figure}

\begin{minipage}{0.43\linewidth}

\centering{

\includegraphics{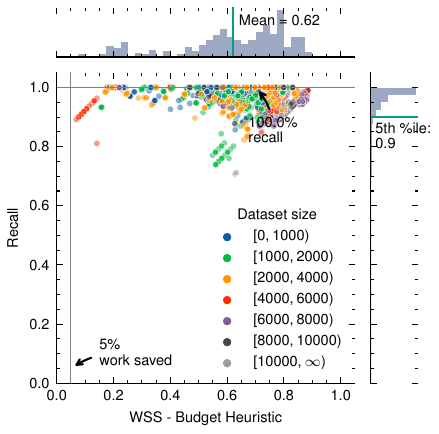}

}

\subcaption{\label{fig-cmhplots-standoff-1}Budget}

\end{minipage}%
\begin{minipage}{0.14\linewidth}
~\end{minipage}%
\begin{minipage}{0.43\linewidth}

\centering{

\includegraphics{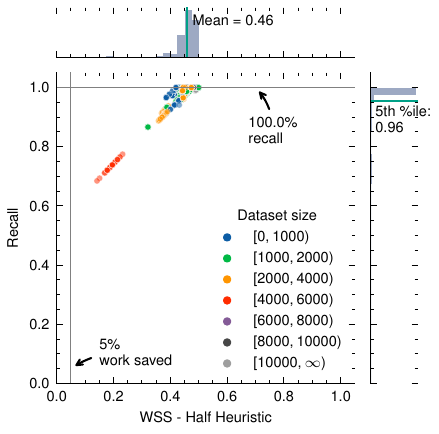}

}

\subcaption{\label{fig-cmhplots-standoff-2}Half}

\end{minipage}%
\newline
\begin{minipage}{0.43\linewidth}

\centering{

\includegraphics{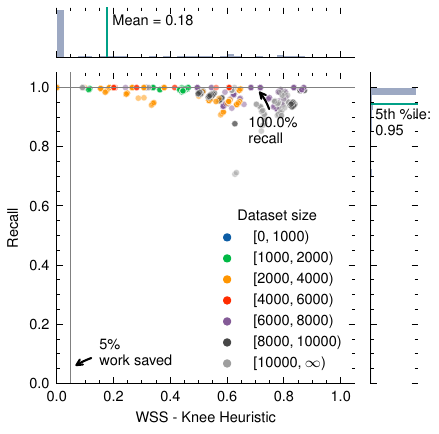}

}

\subcaption{\label{fig-cmhplots-standoff-3}Knee}

\end{minipage}%
\begin{minipage}{0.14\linewidth}
~\end{minipage}%
\begin{minipage}{0.43\linewidth}

\centering{

\includegraphics{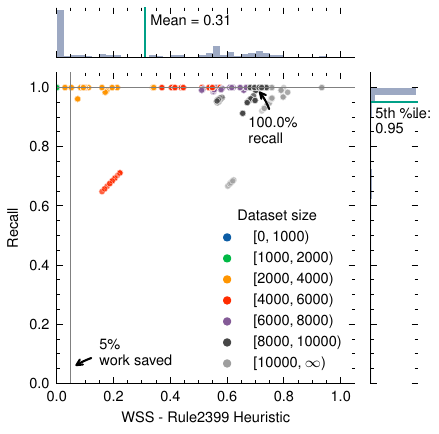}

}

\subcaption{\label{fig-cmhplots-standoff-4}Rule2399}

\end{minipage}%
\newline
\begin{minipage}{0.43\linewidth}

\centering{

\includegraphics{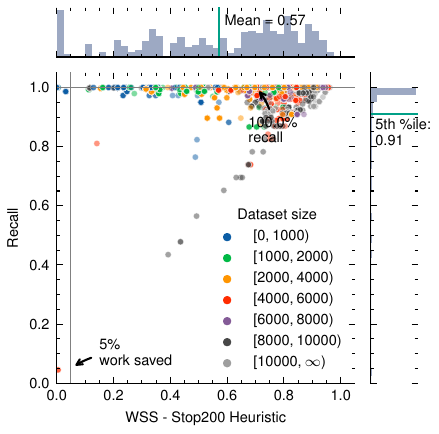}

}

\subcaption{\label{fig-cmhplots-standoff-5}Stop after 200 irrelevant}

\end{minipage}%
\begin{minipage}{0.14\linewidth}
~\end{minipage}%
\begin{minipage}{0.43\linewidth}

\centering{

\includegraphics{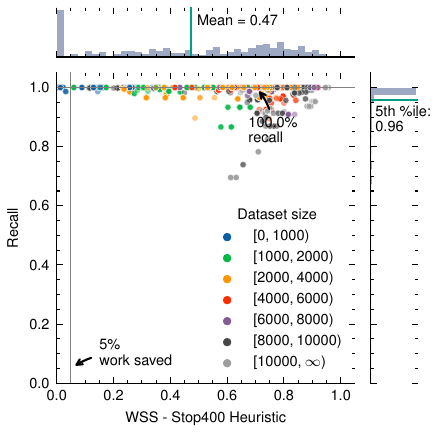}

}

\subcaption{\label{fig-cmhplots-standoff-6}Stop after 400 irrelevant}

\end{minipage}%

\caption{\label{fig-cmhplots-standoff}These figures display the results
of all runs on all datasets in terms of Work Saved over Sampling and
Recall for the standoff heuristic stopping criteria.}

\end{figure}%

The standoff criteria presented in
Table~\ref{tbl-est-comparison-standoff} do not allow specifying recall
targets. All heuristics seem to achieve a high recall; however, some
methods are not always triggered. Especially the Knee heuristic, which
is only triggered \(33.72 %
\) of the time. We suspect that this is due to the fact that there are
many datasets for which the number of relevant documents is below 100,
in which case the Knee has problems {[}15{]}. The Budget method, which
is the best-performing adaptive heuristic, is more adjusted to this. The
(mostly) static heuristics \emph{Half} and \emph{Rule2399} heuristic
also offer good results in terms of recall but are not efficient on
which AutoTAR can provide a good ranking. Moreover, for datasets
containing less than 2399 documents, \emph{Rule2399} does not trigger.
In Table~\ref{tbl-est-comparison-standoff}, the success rates of
achieving various recall targets are presented. Our experiments show
that the Budget method achieves the 95 \% target in 76.25 \% of the
experiments (for \emph{Chao (1987) - Conservative}, this is 69.86 \%).
For the 95 \% recall level, our method does not outperform the Budget
method in terms of recall and work savings. However, for the 100 \%
recall level, it does (Budget 40.63 \% success rate vs.~\emph{Chao
(1987) - Conservative} 53.09 \%).

\bookmarksetup{startatroot}

\section{Discussion}\label{discussion}

In the previous section, we described the results of our simulation
experiment. In this section, we discuss the results and answer the
research questions as posed in Section~\ref{sec-research-questions}.

\subsection{Research Questions}\label{research-questions}

\subsubsection{How does our Active Learning strategy perform in terms of
the WSS@95 and WSS@100 metrics compared to other
methods?}\label{how-does-our-active-learning-strategy-perform-in-terms-of-the-wss95-and-wss100-metrics-compared-to-other-methods}

Here, we compare the retrieval capabilities of the three sampling
strategies used by the stopping criteria we included in our experiments:
AutoTAR, AUTOSTOP, and our Ensemble method. Our method outperforms
AUTOSTOP's sampling strategy but does not outperform AutoTAR in terms of
WSS@95 and WSS@100 (see Table~\ref{tbl-work-comparison}). One can expect
this result as our method performs extra work to enable the use of
Population Size Estimation methods to determine the number of relevant
documents within the dataset. Moreover, our method selects approximately
20 \% of the documents through random sampling, which is not an
efficient retrieval strategy. Despite this, our method stays closer to
AutoTAR's performance, which has a mean WSS@95 score of 74.3 \% vs. ours
(Ensemble) of 63 \% and 45.5 \% for AUTOSTOP. The overhead introduced by
our method is less than AUTOSTOP's, enabling a reduction of the
reviewers' effort given a good stopping criterion.

\subsubsection{Can our stopping criteria help the user achieve its
recall target in a timely
fashion?}\label{can-our-stopping-criteria-help-the-user-achieve-its-recall-target-in-a-timely-fashion}

The \emph{Chao (1987) - Conservative} method with a recall target of 95
\% has a mean recall of 95.92 \% with a Work Saved over Sampling of
56.21 \%. For the 100 \% target, the \emph{Chao (1987) - Conservative}
criterion has a mean recall of 98.02 \%, which is slightly below the
target recall. The \emph{Chao (Rivest) - Optimistic - 100 \%} criterion
achieves an average recall of 98.29 \% with a WSS of 52.33 \%. Although
these results do not always deliver a perfect recall of 100 \%, the
additional burden of using our stopping criterion is not large. When we
compare that \emph{effort}, the percentage of the dataset the reviewer
read, that was performed up to the point the criterion was triggered
(45.96 \%) with the effort required for a perfect stopping criterion
(apriori knowledge; same selection strategy) 51.97 \%, we see that on
average the difference is small; this means that when the criterion
would have been perfect, a similar effort would have been required.
Compared to the most efficient sampling method, AutoTAR, (42.42 \%), our
method would require 3.53 percentage points more work than the most
efficient stopping point.

\subsubsection{How reliable are our stopping
criteria?}\label{how-reliable-are-our-stopping-criteria}

Here, we consider the 95 \% and 100 \% criteria only. While
the~\emph{Chao (1987) - Conservative - 95 \%}~method has a very high
mean recall, which is slightly above its target (95.92 \%), the amount
of times the target has been met is 69.86 \%. However, the mean error in
predicting the recall is small (3.88 points), so when the target is not
achieved, the result is often close to its target. The 100 \% recall
target is achieved in 53.09 \% of the runs for \emph{Chao (1987) -
Conservative - 100 \%}. The Optimistic methods are less reliable than
the Conservative methods due to lacking a CI. Yet, the \emph{Chao
(Rivest) - Optimistic - 100 \%} method provides excellent results, with
a mean recall of 98.29. For all targets, the \emph{Chao (Rivest)}
version of this criterion provides better results than the \emph{Chao
(1987)} variant. Considering point estimates only, the Rivest method is
the obvious choice.

\subsubsection{How do our stopping criteria compare to other methods
that estimate the size of relevant
documents?}\label{how-do-our-stopping-criteria-compare-to-other-methods-that-estimate-the-size-of-relevant-documents}

Considering the Conservative criteria, our methods outperform AUTOSTOP
in terms of WSS and Effort and loss\(_\textrm{er}\), while retaining a
similar recall, although slightly lower. However, on average, the 95 \%
criterion, reaches a recall above its target. As we can see in
Table~\ref{tbl-est-comparison-cons}, the number of times the 95 \%
target is met is higher for AutoSTOP. However, when considering the
Optimistic criteria (Table~\ref{tbl-est-comparison-opt}), the \emph{Chao
(Rivest)} method outperforms AutoSTOP for both the 95 \% and the 100 \%
targets in terms of Work Savings/Effort, Recall, and Reliability.
Compared to the \emph{Chao (1987) - Optimistic}, \emph{Chao (1987) -
Conservative} methods with a recall target of 100 \% also outperform
their AUTOSTOP counterparts in terms of WSS and Effort while providing a
similar recall. Moreover, for the \emph{Chao (1987) - Conservative}
method, our method is more functional compared to AUTOSTOP's
counterpart, as our method is triggered 99.37 \% of the time, vs. 55.93
\%.

The Quant method {[}41{]} overestimates the number of relevant documents
with a large number, which results in the fact that this method is not
often triggered on time (for the 70 \% recall target, the mean recall is
already 98.27 \%). Considering the Conservative Criterion, the situation
worsens due to the overestimation, resulting in this method never being
triggered with a high recall target.

\subsubsection{How do our stopping criteria compare to other methods
that do not provide an
estimate?}\label{how-do-our-stopping-criteria-compare-to-other-methods-that-do-not-provide-an-estimate}

The main other criteria in our experiments are the Knee Method {[}15{]},
Budget Method {[}15{]}, Target Method {[}15{]} and the CMH Method
{[}8{]}. From these methods, the Budget method is very close in terms of
recall, outperforming our method in terms of WSS for the 100 \% recall
target. Considering reliability for the 100 \% target, the \emph{Chao
(1987) - Conservative} method is a better choice, as the success rate is
higher. The Budget method also lacks an estimate for the current level
of recall, which does not give the user any information. The CMH method
allows the specification of a recall target, yet it suffers a similar
fate as the Quant method, as it overestimates the current level of
recall.

\subsection{Limitations}\label{limitations}

There are some limitations to the generalizability of our results
resulting from how we designed our experiments. We will describe these
below.

\subsubsection{Dataset selection}\label{dataset-selection}

We selected datasets for which the number of relevant documents was at
least ten. The main reason for this selection is that our method needs
at least five relevant documents as seed data. Moreover, the Target and
Budget methods expect at least ten relevant documents in the dataset. As
all methods are initialized with seed sets that contain five relevant
documents, we deemed that including datasets that contain less than ten
documents would provide unrealistic results. The consequence is that we
cannot extend our findings to all datasets in general; however, the vast
majority of the datasets/topics within the corpora included in our study
contained more than ten relevant documents. Moreover, we excluded
dataset sets with less than 500 documents; however, the merits using TAR
in datasets that small is low.

\subsubsection{Size and contents of the seed
set}\label{size-and-contents-of-the-seed-set}

As mentioned, our methods need a seed set that contains five relevant
documents. The choice of seed documents may influence the results of our
method. In our experiments, we aimed to control for the effect of the
seed set by repeating the experiment 30 times. In each run, a different
seed set was used by using 30 random samples. In a real-world
application, the set of seed documents the user provides may contain
relevant documents that were selected~\emph{not}~at random (for example,
documents with very similar content). The 30 seed sets are likely not
exhaustive enough to capture all these scenarios. Further investigation
is needed to control for these scenarios, for example, by increasing the
number of experiments or testing for specific seed sets.

\subsubsection{Strictness of the Conservative
Criterion}\label{strictness-of-the-conservative-criterion}

In the Conservative Criterion, the recall is calculated using the upper
bound of the confidence interval provided by the Estimator. For very
high recall targets (e.g., 95 \% and especially 100 \%), the confidence
interval must be negligible to trigger the criterion. In some scenarios,
for example, in Figure~\ref{fig-recall-plot-2},~the confidence interval
of the \emph{Chao (Rivest)} method is only one unit above the point
estimates at iteration \(t= 8000\), which already coincides with the
recall curve. The dataset in question, the \emph{Van Dis} dataset,
contains 72 relevant documents so the upper bound of the recall estimate
is currently 98.6 \% (given an upper bound of 73), thus smaller than the
100 \% required by the criterion. This strictness is not only present
within our method but also for AUTOSTOP and Quant. This effect is less
pronounced in datasets with much more relevant documents, as a single
document has less influence on the recall estimate. In a real-world
scenario, the user may deviate from the strictness of this criterion and
be more lenient; for example, if the point estimates or CI are stable
for a long time, the user may accept a slightly smaller predicted
recall. It is not trivial to make a stopping criterion that considers
this leniency. We opted to be very strict in our criterion, as making
the stopping criterion more complex also makes interpreting the results
more challenging. In a real-world setting, the results of our methods
and AUTOSTOP may be better (or worse) if the user freely decides to stop
based on the estimates produced by the estimators.

\subsection{Future work}\label{future-work}

While the results of our experiments are promising, some areas can be
further investigated. As mentioned before, the results of the optimistic
criteria and, in turn, the point estimates of the Chao Estimators are
not as reliable as desired. This result was also observed for AUTOSTOP
and the methods the authors tested in {[}31{]}. Moreover, the error of
the estimators is higher for the lower recall targets, also when we
consider the confidence interval. Besides Chao's Estimator, there are
various other Population Size Estimators available (for instance, models
that take dependencies between committee members into account), as well
as extensions to Chao's Estimator (e.g., extending the model with
covariates found in the dataset). In future research, these methods can
be explored and compared to the results presented here.

A second line of work is to perform a user study, which studies how and
when users stop the review given the decision by a stopping criterion or
an estimate. This study could give insights into how users respond to
the predictions by PSE methods, which could be further used to adapt the
stopping criterion.

As mentioned above, the seed set's size and contents may influence the
performance of our method. Moreover, in a real-world setting, the user
must have five relevant documents available to initialize the AL
procedure, which is not always possible. In the scenario where little
prior known relevant documents are available, a system can be designed
that first finds relevant documents with (for example) AutoTAR and
switches to our method when sufficient documents are found. Another
option is to generate several distinct synthetic documents using Large
Language Models to start the procedure in place of real relevant
documents. Research and development of these extensions may improve the
applicability of our method.

\bookmarksetup{startatroot}

\section{Conclusion}\label{conclusion}

In this work, we used Chao's Population Size Estimator to determine the
number of relevant documents during the TAR process. This estimate can
indicate if the number of documents that have yet eluded the reviewers.
The reviewers can then, given this information, decide to stop the
review when a recall target has been satisfied. Population Size
Estimators are not directly applicable to the general CAL paradigm. In
this work, we presented a novel sampling strategy that makes these
methods possible while minimizing the number of irrelevant documents the
system proposes. We employed two versions of Chao's estimator,
\emph{Chao (1987)}, based on Chao's Moment estimator as presented in
{[}10{]}. The other, \emph{Chao (Rivest)} is based on Rivest's Poisson
Regression version of the former as presented in {[}32{]}. The estimates
from these methods are then used within a stopping criterion for the
review process. For each estimator, we built two criteria: an Optimistic
method, which uses the point estimates, while the Conservative uses the
95 \% confidence interval. An extensive simulation study showed us that
the proposed estimators and criteria work well. The \emph{Chao (Rivest)
- Optimistic} method clearly outperforms other estimator-based methods
regarding recall and work savings. The \emph{Chao (1987) - Conservative}
method challenges methods presented in previous work, as it achieves a
similar level recall while improving work savings. We expect that
further research into PSE and extensions of our method will improve the
reliability and applicability of this method.

\bookmarksetup{startatroot}

\section*{References}\label{references}
\addcontentsline{toc}{section}{References}

\markboth{References}{References}

\phantomsection\label{refs}

\cleardoublepage
\phantomsection
\addcontentsline{toc}{part}{Appendices}
\appendix

\section{Dataset statistics}\label{sec-dataset-statistics-for-slr}

\begin{longtable}[t]{lrrrr}

\caption{\label{tbl-datasets-slr}Dataset statistics for the SYNERGY
corpus {[}18{]}}

\tabularnewline

\toprule
\textbf{Dataset} & \textbf{\# Relevant} & \textbf{\# Irrelevant} & \textbf{Size} & \textbf{Prevalence (\%)}\\
\midrule
\cellcolor{gray!15}{Appenzeller-Herzog} & \cellcolor{gray!15}{26} & \cellcolor{gray!15}{2847} & \cellcolor{gray!15}{2873} & \cellcolor{gray!15}{0.9}\\
Brouwer & 62 & 38052 & 38114 & 0.2\\
\cellcolor{gray!15}{Chou} & \cellcolor{gray!15}{15} & \cellcolor{gray!15}{1893} & \cellcolor{gray!15}{1908} & \cellcolor{gray!15}{0.8}\\
Hall & 104 & 8689 & 8793 & 1.2\\
\cellcolor{gray!15}{Jeyaraman} & \cellcolor{gray!15}{96} & \cellcolor{gray!15}{1079} & \cellcolor{gray!15}{1175} & \cellcolor{gray!15}{8.2}\\
\addlinespace
Leenaars & 583 & 6633 & 7216 & 8.1\\
\cellcolor{gray!15}{Meijboom} & \cellcolor{gray!15}{37} & \cellcolor{gray!15}{845} & \cellcolor{gray!15}{882} & \cellcolor{gray!15}{4.2}\\
Menon & 74 & 901 & 975 & 7.6\\
\cellcolor{gray!15}{Moran} & \cellcolor{gray!15}{111} & \cellcolor{gray!15}{5103} & \cellcolor{gray!15}{5214} & \cellcolor{gray!15}{2.1}\\
Muthu & 336 & 2383 & 2719 & 12.4\\
\addlinespace
\cellcolor{gray!15}{Oud} & \cellcolor{gray!15}{20} & \cellcolor{gray!15}{932} & \cellcolor{gray!15}{952} & \cellcolor{gray!15}{2.1}\\
Radjenovic & 48 & 5887 & 5935 & 0.8\\
\cellcolor{gray!15}{Smid} & \cellcolor{gray!15}{27} & \cellcolor{gray!15}{2600} & \cellcolor{gray!15}{2627} & \cellcolor{gray!15}{1.0}\\
Walker & 762 & 47613 & 48375 & 1.6\\
\cellcolor{gray!15}{Wassenaar} & \cellcolor{gray!15}{111} & \cellcolor{gray!15}{7557} & \cellcolor{gray!15}{7668} & \cellcolor{gray!15}{1.4}\\
\addlinespace
Wolters & 19 & 4261 & 4280 & 0.4\\
\cellcolor{gray!15}{van Dis} & \cellcolor{gray!15}{72} & \cellcolor{gray!15}{9056} & \cellcolor{gray!15}{9128} & \cellcolor{gray!15}{0.8}\\
van de Schoot & 38 & 4506 & 4544 & 0.8\\
\cellcolor{gray!15}{van der Valk} & \cellcolor{gray!15}{89} & \cellcolor{gray!15}{636} & \cellcolor{gray!15}{725} & \cellcolor{gray!15}{12.3}\\
van der Waal & 33 & 1937 & 1970 & 1.7\\
\bottomrule

\end{longtable}

\begin{longtable}[t]{lrrrr}

\caption{\label{tbl-datasets-clef2017}Dataset statistics for the CLEF
2017 corpus {[}24{]}}

\tabularnewline

\toprule
\textbf{Dataset} & \textbf{\# Relevant} & \textbf{\# Irrelevant} & \textbf{Size} & \textbf{Prevalence (\%)}\\
\midrule
\endfirsthead
\multicolumn{5}{@{}l}{\textit{(continued)}}\\
\toprule
\textbf{Dataset} & \textbf{\# Relevant} & \textbf{\# Irrelevant} & \textbf{Size} & \textbf{Prevalence (\%)}\\
\midrule
\endhead

\endfoot
\bottomrule
\endlastfoot
\cellcolor{gray!15}{CD009135} & \cellcolor{gray!15}{77} & \cellcolor{gray!15}{714} & \cellcolor{gray!15}{791} & \cellcolor{gray!15}{9.7}\\
CD008081 & 26 & 944 & 970 & 2.7\\
\cellcolor{gray!15}{CD010023} & \cellcolor{gray!15}{52} & \cellcolor{gray!15}{929} & \cellcolor{gray!15}{981} & \cellcolor{gray!15}{5.3}\\
CD009944 & 98 & 1064 & 1162 & 8.4\\
\cellcolor{gray!15}{CD008691} & \cellcolor{gray!15}{67} & \cellcolor{gray!15}{1243} & \cellcolor{gray!15}{1310} & \cellcolor{gray!15}{5.1}\\
\addlinespace
CD007427 & 59 & 1398 & 1457 & 4.0\\
\cellcolor{gray!15}{CD010632} & \cellcolor{gray!15}{27} & \cellcolor{gray!15}{1472} & \cellcolor{gray!15}{1499} & \cellcolor{gray!15}{1.8}\\
CD009020 & 154 & 1422 & 1576 & 9.8\\
\cellcolor{gray!15}{CD009185} & \cellcolor{gray!15}{92} & \cellcolor{gray!15}{1523} & \cellcolor{gray!15}{1615} & \cellcolor{gray!15}{5.7}\\
CD009551 & 46 & 1865 & 1911 & 2.4\\
\addlinespace
\cellcolor{gray!15}{CD011134} & \cellcolor{gray!15}{200} & \cellcolor{gray!15}{1738} & \cellcolor{gray!15}{1938} & \cellcolor{gray!15}{10.3}\\
CD009372 & 25 & 2223 & 2248 & 1.1\\
\cellcolor{gray!15}{CD007394} & \cellcolor{gray!15}{92} & \cellcolor{gray!15}{2450} & \cellcolor{gray!15}{2542} & \cellcolor{gray!15}{3.6}\\
CD009647 & 56 & 2729 & 2785 & 2.0\\
\cellcolor{gray!15}{CD008054} & \cellcolor{gray!15}{206} & \cellcolor{gray!15}{2940} & \cellcolor{gray!15}{3146} & \cellcolor{gray!15}{6.5}\\
\addlinespace
CD010438 & 30 & 3211 & 3241 & 0.9\\
\cellcolor{gray!15}{CD009323} & \cellcolor{gray!15}{98} & \cellcolor{gray!15}{3757} & \cellcolor{gray!15}{3855} & \cellcolor{gray!15}{2.5}\\
CD008803 & 99 & 5121 & 5220 & 1.9\\
\cellcolor{gray!15}{CD010173} & \cellcolor{gray!15}{23} & \cellcolor{gray!15}{5472} & \cellcolor{gray!15}{5495} & \cellcolor{gray!15}{0.4}\\
CD010276 & 54 & 5441 & 5495 & 1.0\\
\addlinespace
\cellcolor{gray!15}{CD009519} & \cellcolor{gray!15}{104} & \cellcolor{gray!15}{5867} & \cellcolor{gray!15}{5971} & \cellcolor{gray!15}{1.7}\\
CD009579 & 138 & 6317 & 6455 & 2.1\\
\cellcolor{gray!15}{CD009925} & \cellcolor{gray!15}{460} & \cellcolor{gray!15}{6071} & \cellcolor{gray!15}{6531} & \cellcolor{gray!15}{7.0}\\
CD009591 & 143 & 7847 & 7990 & 1.8\\
\cellcolor{gray!15}{CD010653} & \cellcolor{gray!15}{45} & \cellcolor{gray!15}{7957} & \cellcolor{gray!15}{8002} & \cellcolor{gray!15}{0.6}\\
\addlinespace
CD011984 & 442 & 7738 & 8180 & 5.4\\
\cellcolor{gray!15}{CD011975} & \cellcolor{gray!15}{604} & \cellcolor{gray!15}{7582} & \cellcolor{gray!15}{8186} & \cellcolor{gray!15}{7.4}\\
CD008782 & 45 & 10462 & 10507 & 0.4\\
\cellcolor{gray!15}{CD011548} & \cellcolor{gray!15}{109} & \cellcolor{gray!15}{12591} & \cellcolor{gray!15}{12700} & \cellcolor{gray!15}{0.9}\\
CD010339 & 114 & 12689 & 12803 & 0.9\\
\addlinespace
\cellcolor{gray!15}{CD009593} & \cellcolor{gray!15}{63} & \cellcolor{gray!15}{14844} & \cellcolor{gray!15}{14907} & \cellcolor{gray!15}{0.4}\\*

\end{longtable}

\begin{longtable}[t]{lrrrr}

\caption{\label{tbl-datasets-clef2018}Dataset statistics for the CLEF
2018 corpus {[}25{]}}

\tabularnewline

\toprule
\textbf{Dataset} & \textbf{\# Relevant} & \textbf{\# Irrelevant} & \textbf{Size} & \textbf{Prevalence (\%)}\\
\midrule
\endfirsthead
\multicolumn{5}{@{}l}{\textit{(continued)}}\\
\toprule
\textbf{Dataset} & \textbf{\# Relevant} & \textbf{\# Irrelevant} & \textbf{Size} & \textbf{Prevalence (\%)}\\
\midrule
\endhead

\endfoot
\bottomrule
\endlastfoot
\cellcolor{gray!15}{CD012009} & \cellcolor{gray!15}{37} & \cellcolor{gray!15}{499} & \cellcolor{gray!15}{536} & \cellcolor{gray!15}{6.9}\\
CD008759 & 60 & 872 & 932 & 6.4\\
\cellcolor{gray!15}{CD011431} & \cellcolor{gray!15}{297} & \cellcolor{gray!15}{885} & \cellcolor{gray!15}{1182} & \cellcolor{gray!15}{25.1}\\
CD011912 & 36 & 1370 & 1406 & 2.6\\
\cellcolor{gray!15}{CD008892} & \cellcolor{gray!15}{69} & \cellcolor{gray!15}{1430} & \cellcolor{gray!15}{1499} & \cellcolor{gray!15}{4.6}\\
\addlinespace
CD010657 & 139 & 1720 & 1859 & 7.5\\
\cellcolor{gray!15}{CD008122} & \cellcolor{gray!15}{272} & \cellcolor{gray!15}{1639} & \cellcolor{gray!15}{1911} & \cellcolor{gray!15}{14.2}\\
CD011053 & 12 & 2223 & 2235 & 0.5\\
\cellcolor{gray!15}{CD010864} & \cellcolor{gray!15}{44} & \cellcolor{gray!15}{2461} & \cellcolor{gray!15}{2505} & \cellcolor{gray!15}{1.8}\\
CD010502 & 229 & 2756 & 2985 & 7.7\\
\addlinespace
\cellcolor{gray!15}{CD011926} & \cellcolor{gray!15}{40} & \cellcolor{gray!15}{4010} & \cellcolor{gray!15}{4050} & \cellcolor{gray!15}{1.0}\\
CD010296 & 53 & 4549 & 4602 & 1.2\\
\cellcolor{gray!15}{CD009175} & \cellcolor{gray!15}{65} & \cellcolor{gray!15}{5579} & \cellcolor{gray!15}{5644} & \cellcolor{gray!15}{1.2}\\
CD011126 & 13 & 5987 & 6000 & 0.2\\
\cellcolor{gray!15}{CD012010} & \cellcolor{gray!15}{290} & \cellcolor{gray!15}{6540} & \cellcolor{gray!15}{6830} & \cellcolor{gray!15}{4.2}\\
\addlinespace
CD011515 & 127 & 7117 & 7244 & 1.8\\
\cellcolor{gray!15}{CD012599} & \cellcolor{gray!15}{575} & \cellcolor{gray!15}{7473} & \cellcolor{gray!15}{8048} & \cellcolor{gray!15}{7.1}\\
CD010680 & 26 & 8379 & 8405 & 0.3\\
\cellcolor{gray!15}{CD008587} & \cellcolor{gray!15}{79} & \cellcolor{gray!15}{9073} & \cellcolor{gray!15}{9152} & \cellcolor{gray!15}{0.9}\\
CD011686 & 64 & 9665 & 9729 & 0.7\\
\addlinespace
\cellcolor{gray!15}{CD012179} & \cellcolor{gray!15}{304} & \cellcolor{gray!15}{9528} & \cellcolor{gray!15}{9832} & \cellcolor{gray!15}{3.1}\\
CD012281 & 23 & 9853 & 9876 & 0.2\\
\cellcolor{gray!15}{CD012165} & \cellcolor{gray!15}{308} & \cellcolor{gray!15}{9914} & \cellcolor{gray!15}{10222} & \cellcolor{gray!15}{3.0}\\
CD010213 & 599 & 14599 & 15198 & 3.9\\*

\end{longtable}

\begin{longtable}[t]{lrrrr}

\caption{\label{tbl-datasets-clef2019}Dataset statistics for the CLEF
2019 corpus {[}26{]}}

\tabularnewline

\toprule
\textbf{Dataset} & \textbf{\# Relevant} & \textbf{\# Irrelevant} & \textbf{Size} & \textbf{Prevalence (\%)}\\
\midrule
\endfirsthead
\multicolumn{5}{@{}l}{\textit{(continued)}}\\
\toprule
\textbf{Dataset} & \textbf{\# Relevant} & \textbf{\# Irrelevant} & \textbf{Size} & \textbf{Prevalence (\%)}\\
\midrule
\endhead

\endfoot
\bottomrule
\endlastfoot
\cellcolor{gray!15}{CD001261} & \cellcolor{gray!15}{72} & \cellcolor{gray!15}{499} & \cellcolor{gray!15}{571} & \cellcolor{gray!15}{12.6}\\
CD012551 & 68 & 523 & 591 & 11.5\\
\cellcolor{gray!15}{CD007867} & \cellcolor{gray!15}{17} & \cellcolor{gray!15}{926} & \cellcolor{gray!15}{943} & \cellcolor{gray!15}{1.8}\\
CD012669 & 71 & 1189 & 1260 & 5.6\\
\cellcolor{gray!15}{CD009069} & \cellcolor{gray!15}{78} & \cellcolor{gray!15}{1679} & \cellcolor{gray!15}{1757} & \cellcolor{gray!15}{4.4}\\
\addlinespace
CD009642 & 62 & 1860 & 1922 & 3.2\\
\cellcolor{gray!15}{CD008874} & \cellcolor{gray!15}{118} & \cellcolor{gray!15}{2264} & \cellcolor{gray!15}{2382} & \cellcolor{gray!15}{5.0}\\
CD010753 & 29 & 2510 & 2539 & 1.1\\
\cellcolor{gray!15}{CD010558} & \cellcolor{gray!15}{37} & \cellcolor{gray!15}{2778} & \cellcolor{gray!15}{2815} & \cellcolor{gray!15}{1.3}\\
CD009044 & 11 & 3158 & 3169 & 0.3\\
\addlinespace
\cellcolor{gray!15}{CD012661} & \cellcolor{gray!15}{192} & \cellcolor{gray!15}{3175} & \cellcolor{gray!15}{3367} & \cellcolor{gray!15}{5.7}\\
CD012069 & 320 & 3159 & 3479 & 9.2\\
\cellcolor{gray!15}{CD006468} & \cellcolor{gray!15}{52} & \cellcolor{gray!15}{3821} & \cellcolor{gray!15}{3873} & \cellcolor{gray!15}{1.3}\\
CD011787 & 111 & 4258 & 4369 & 2.5\\
\cellcolor{gray!15}{CD012080} & \cellcolor{gray!15}{77} & \cellcolor{gray!15}{6566} & \cellcolor{gray!15}{6643} & \cellcolor{gray!15}{1.2}\\
\addlinespace
CD012567 & 11 & 6724 & 6735 & 0.2\\
\cellcolor{gray!15}{CD010038} & \cellcolor{gray!15}{23} & \cellcolor{gray!15}{8844} & \cellcolor{gray!15}{8867} & \cellcolor{gray!15}{0.3}\\
CD011768 & 54 & 9104 & 9158 & 0.6\\
\cellcolor{gray!15}{CD011686} & \cellcolor{gray!15}{64} & \cellcolor{gray!15}{9665} & \cellcolor{gray!15}{9729} & \cellcolor{gray!15}{0.7}\\*

\end{longtable}

\end{document}